\documentclass[useAMS,usenatbib,usegraphicx]{mn2e}
 
\title[Galaxy Structures in the UDF]{The Structures of Distant Galaxies I:  Galaxy Structures and the Merger Rate to $z \sim 3$ in the Hubble Ultra-Deep Field}
\author[Conselice, Rajgor, Myers]{ Christopher J. Conselice$^{1}$\thanks{E-mail:
conselice@nottingham.ac.uk}, Sheena Rajgor$^{1}$, Robert Myers$^{1}$ \\
$^{1}$University of Nottingham, School of Physics \& Astronomy, Nottingham, NG7 2RD UK }
\def\deg{$^{\circ}\,$}
\def\solm{M$_{\odot}\,$}

\def\deg{$^{\circ}\,$}
\def\solm{M$_{\odot}\,$}

\def\mass{$10^{11}$ M$_{\odot}\,$}

\def\lmass{$10^{10}$ M$_{\odot}\,$}
\def\llmass{$10^{9}$ M$_{\odot}\,$}
\def\lllmass{$10^{8}$ M$_{\odot}\,$}
\def\casgm20{CAS-G-M$_{20}\,$}
\def\m20{M$_{20}\,$}

\begin{document}

\date{Accepted ; Received ; in original form}

\pagerange{\pageref{firstpage}--\pageref{lastpage}} \pubyear{2002}

\maketitle

\label{firstpage}

\begin{abstract}

This paper begins a series in which we examine the structures of distant
galaxies to directly determine the history of their formation modes.  
We start this series by examining the structures of 
$z_{F850LP} < 27$ galaxies in the Hubble Ultra-Deep field, the deepest 
high-resolution optical image taken to date.  We investigate a few basic 
features of galaxy structure using this image.  These include: (1) 
The agreement 
of visual eye-ball classifications and non-parametric quantitative 
(CAS, Gini/\m20) methods;  (2) How distant galaxy quantitative
structures can vary as a function of rest-frame wavelength; and (3) 
The evolution of distant galaxy structures up to $z \sim 3$.  One of 
our major conclusions is that the majority of galaxies with $z_{850} < 27$ 
are peculiar in appearance, and
that galaxy assembly is rapidly occurring at these magnitudes, even
up to the present time.   We find a general agreement
between galaxy classification by eye and through quantitative methods, as
well as a general agreement between the CAS and the Gini/\m20 parameters.  We
find that the Gini/\m20 method appears to find a larger number of galaxy 
mergers than the CAS system, but contains a larger contamination from 
non-mergers. We furthermore calculate the merger rate of 
galaxies in the UDF up to $z \sim 3$, finding an increase with redshift as 
well as stellar mass, confirming previous work in the Hubble Deep Field.   
We find that massive galaxies with M$_{*} >$ \lmass undergo
4.3$_{+0.8}^{-0.8}$ major galaxy mergers at $z < 3$, with all of this merging
occurring at $z > 1$.

\end{abstract}


\section{Introduction}

Understanding the formation of galaxies is one of the most intriguing
and unanswered problems in astronomy today. Studying galaxies
in the distant universe is one of the major approaches for deciphering
their formation and evolution.  Traditionally, distant galaxies
are examined in terms of their gross properties, such as
luminosities (e.g., Faber et al. 2007), colours (e.g.,
Moustakas et al. 2004), and sizes (Trujillo et al.
2007), and stellar masses (e.g., Bundy et al. 2005; Conselice
et al. 2007a,b).  The least developed
approached, but potentially one of the most exciting for revealing unique
information, is how the structures
and morphologies of galaxies change through cosmic time.

The reason a structural and morphological approach is potentially the 
most fruitful for understanding galaxy formation and evolution is that 
the structures of galaxies correlate with formation modes, such as 
star formation and galaxy mergers (Conselice 2003).  It is thus possible to
determine the physical processes behind galaxy formation directly
without having to compare observables with physical models.
This direct approach has been used in the past to calculate the role
of galaxy mergers in the formation of massive galaxies (e.g.,
Conselice et al. 2003a; Lotz et al. 2006; Conselice 2006).  This previous 
work showed that the most massive galaxies with M$_{*} >$ \lmass appear
to form nearly all their stellar mass in mergers, which occur mostly,
and in large numbers, at $z > 1.5$.

There are however still many outstanding problems with using galaxy structure
to derive evolution, including the reliability of the approaches used,
and the uniqueness and repeatability of results thus far obtained.  
For the most part, it is necessary to use deep Hubble Space Telescope 
imaging to measure the structures and morphologies of a large number
of distant galaxies, although adaptive
optics, particularly in the near-infrared, is becoming another powerful
approach (e.g., Wright et al. 2007).  Since the imaging cameras
on the Hubble Space Telescope are limited in field of view, only
a small fraction of distant galaxies have been studied thus
far (e.g., Williams et al. 1996; Giavalisco et al. 2004). Particularly,
most of the merger history of galaxies at $z > 1$ has been measured
within the Hubble Deep Field-North, which currently has the only
high resolution deep near-infrared imaging of distant galaxies
(e.g., Dickinson et al. 2000; Conselice et al. 2003a; Papovich et al.
2005).  Thus, it is important to study the structures
and morphologies of galaxies in as many fields as possible, using different
techniques.

The Hubble Ultra Deep Field (UDF) represents perhaps our best
opportunity to study the most distant galaxies at observed optical wavelengths
for the foreseeable future. The UDF is a major Hubble imaging program
utilising the Advanced Camera for Surveys (ACS) to image a single
pointing in the sky to the deepest depth ever probed in the 
optical.  As the ACS UDF images are very deep, and have a higher resolution 
than the WFPC2 HDFs, we can address some fundamental questions concerning
galaxy structure and morphology.  These include:
understanding the nature of galaxy morphologies
seen in deep optical surveys; the agreement, or lack thereof, between
apparent morphology and structural parameters; and the evolution of
galaxy structure through time.  We are particularly interested in 
determine how major galaxy mergers are driving the formation of
the galaxy population at early times.

Our major finding in this paper is that galaxies in the past are 
significantly more
irregular, peculiar, asymmetric, and clumpy than galaxies today, and that 
this likely results from the past merger and assembly 
activity that established these systems.   We show that a large
fraction of this irregularity cannot be produced simply by
star formation, and derive the merger fraction and rate for galaxies
of various stellar masses.
We confirm the findings of Conselice et al. (2003a) and Conselice
(2006) that massive galaxies with M$_{*} >$ \lmass have a steeply
increasing merger fraction, which evolves as $\sim$ $(1+z)^{6}$ between
$z = 0.8$ and 3.   This suggests that these massive galaxies undergo
on average 4.3$^{+0.8}_{-0.8}$ major mergers between $z \sim 3$ to 0.

This paper is organised as follows: \S 2 includes a discussion of the 
data sources we use in this paper, \S 3 is a description of our 
morphological and structural analyses, and the stellar masses we utilise, 
\S 4 is a discussion of our
results, including how different structural analysis techniques compare
with each other, especially how merger finding methods differ, \S 5 gives
our description of the merger history up to $z \sim 3$, and \S 6 is our 
summary and conclusions.   We use a standard cosmology
of H$_{0} = 70$ km s$^{-1}$ Mpc$^{-1}$, and 
$\Omega_{\rm m} = 1 - \Omega_{\lambda}$ = 0.3 throughout.

\section{Data and Data Sources}

The primary data source used in this paper are the ACS and NICMOS
imaging of the Hubble Ultra Deep Field (Thompson et al. 2005;
Beckwith et al. 2006).  The field of view of the ACS image is 11 
arcmin$^{2}$, and is located within the GOODS-South field 
(Giavalisco et al. 2004).  The UDF ACS images use the same filter 
set as the GOODS data, which are the F435W (B$_{435}$), F606W 
(V$_{606}$), F775W ($i_{775}$), and F850L ($z_{815}$) bands.   The central
wavelengths of these filters, and their full-width at half-maximum
are: F435W (4297$\pm1038$ \AA), F606W (5907$\pm2342$ \AA), 
F775W (7764$\pm1528$ \AA), F850L (9445$\pm1229$ \AA). The limiting magnitude
for point sources is m$_{\rm AB} \sim 29$ within these images, making
the UDF easily the deepest optical imaging taken to date.

The photometry and photometric redshifts we use are taken from Coe et al. 
(2006).    Coe et al. (2006) measure the photometry of galaxies in the UDF 
within the UDF BV$iz$JH bands with great care.  The galaxies are detected 
with a modified version of SExtractor, called SExSeg. The
photometry is PSF-corrected and aperture matched, removing the
problem of matching magnitudes at different wavelengths due to 
variations in the PSF.  The $z_{850}$ data was furthermore corrected 
for PSF halo effects, and a correction was made to the NICMOS photometric 
zero point (Coe et al. 2006).  This high fidelity photometry is then used
for deriving photometric redshifts, and for the stellar masses we measure
for our galaxy sample.
  
The Coe et al. (2006) photometric redshifts are measured using the photometric
redshift techniques from Benitez (2000), based on the optical and NIR 
photometry from ACS and NICMOS. In addition to these photometric 
redshifts we utilise 56 spectroscopic redshifts taken to date
within the UDF field within our selection limits.

We limit our analysis of the UDF galaxies to the relatively brighter 
systems, although
we are able to probe in the UDF down to a fainter magnitude than any other 
field. 
Faint galaxies are difficult to study structurally, as they are often too 
small, and have too low a surface brightness for reliable measurements. 
We thus limit the magnitude of our study to $z_{850} < 27$, such that we are 
not biased by low signal to noise imaging.    We furthermore use the 
Coe et al.(2006)  UDF image versions with their accompanying 
segmentation map from SExtractor, and weight map, which we 
utilize later in our structural analysis (\S 3).  The final 
catalogue of $z_{850} < 27$ sources we study in this paper contains 
1052 unique galaxies. 

\section{Structural Measures and Stellar Masses}

As described in the introduction, the structures and morphologies of galaxies 
is quickly becoming recognised as one of the most important methods for 
understanding galaxies (e.g., Conselice 2003a,b; Cassata et al. 2005; Grogin
et al. 2005;  Trujillo et al. 2007).  As such, we study the  morphological properties 
of all $z_{850} < 27$ galaxies in the UDF in some detail.   It is important 
however to realise what we can, and cannot, study within the UDF.  Although 
the UDF is the deepest image ever taken, it is unfortunately not very 
large, with a total area of 11 arcmin$^{2}$.  This does not give a 
large enough co-moving volume to study galaxies in large detail at any
redshift. It does allow us to probe structures and morphologies
in greater detail than in any other field, although general results
on galaxy evolution obtained from it are limited.

\begin{figure}
 \vbox to 120mm{
\includegraphics[angle=0, width=90mm]{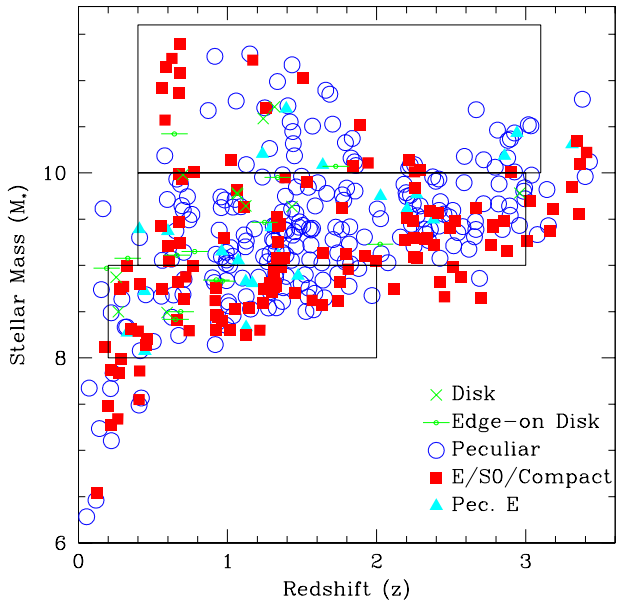}
 \caption{Stellar mass vs. redshift with the various galaxy
types studied in this paper labelled.  The open blue circles are
the galaxies classified as peculiars, the solid red boxes are ellipticals,
S0s, and compacts, while the cyan triangles are ellipticals that appear to
have a peculiarity.  Disk galaxies are shown as green crosses for face
on systems, with the edge-on systems displayed as a dot with a solid
line.  The boxes on this figures denote the regions in which we measure
the merger fraction later in the paper.}
} \label{sample-figure}
\end{figure}

We carry out our analysis in several steps to maximise the usefulness
of the data, and to minimise problems from contamination.
We first create `postage stamp' images of each of our sample galaxies.
These are created by cutting out a 10'' $\times$ 10'' box of the UDF 
surrounding each galaxy, based on positions from the SExtractor
catalog detections from Coe et al. (2006).   Before this is done, the UDF
ACS image is cleaned of nearby galaxies
and stars through the use of the so-called `segmentation map' produced
by the SExSeg (\S 2, Coe et al. 2006). The segmentation maps
are created through the SExtractor procedure used by Coe et al.
(2006) for detecting galaxies within the UDF.  These segmentation maps
are equivalent in size to the UDF image itself, with the difference
being that it gives a numerical value for each pixel that reveals
which galaxy it belongs to.  These segmentation maps are used for
photometry, but they are also useful for removing nearby galaxies.
The procedure we use is to replace pixels of galaxies not being studied
to the sky background with proper noise characteristics included.  We then 
use these cleaned cutout images in our analysis.

After examining our sample by eye, we found that occasionally features
remained near galaxies, and had to be manually removed by hand. There were 
also cases where large late-type galaxies with spiral arms 
brighter than their centres tended to be picked up by the program more than 
once, and
these were manually noted when spotted.  In the following
sections we describe our visual and quantitative analysis of these galaxy
images within the UDF.

\subsection{Visual/Classical Morphologies}

We study the structures and morphologies of our sample using two broadly
different methods.   The first is a simple visual estimate of morphologies 
based on the appearance of our galaxies in the ACS imaging.   The outline of 
our classification process is given in Conselice et al. (2005a), and 
Conselice et al. (2007b).    We place each UDF galaxy into one of nine 
categories: compact, elliptical, distorted elliptical, lenticular (S0),
early-type disk,  late-type disk, edge-on disk, merger/peculiar, 
and unknown/too-faint.  Our classifications are based only on appearance. 
Information such as colour, size, redshift, etc are not used to determine 
these types. We carry out these classifications to  link our results
with lower redshifts, as well as for having some basis for understanding
the morphological distribution of galaxies at high redshifts.   A short 
explanation of these types is provided below, with the number we find in 
each class listed at the end of each description.

\begin{enumerate}

 \item Ellipticals : Ellipticals (Es) are centrally concentrated galaxies with 
no evidence for lower surface brightness, outer structures.  We have
128 of these galaxies in our sample.  

 \item Peculiar-Ellipticals : Peculiar ellipticals  (Pec-Es) are galaxies that
appear elliptical, but have some minor morphological peculiarity,
such as offset isophotes, dual nuclei, or low-surface brightness asymmetries
in their outer parts (65 systems).

  \item S0s: S0s are galaxies that appear to have a smooth disk with
a bulge. These galaxies do not appear to have much star formation,
and are selected in the same way nearby S0s are.  Our sample
has a total of three S0s, making its contribution very small.

 \item Compact - A galaxy is classified as compact if its structure
is resolved, but still appears compact without any substructure. It is 
similar to the elliptical classification in that
a system must appear very smooth and symmetric. A compact galaxy differs
from an elliptical in that it contains no obvious features such as an extended
light distribution or a light envelope.  (153 systems)
     
 \item Early-type disks: If a galaxy contains a central 
concentration with some evidence for lower surface brightness outer 
light in the form of spiral arms or a disk, it is classified as an 
early-type disk. (28 systems)
 
 \item Late-type disks: Late-type disks are galaxies that appear to 
have more outer low surface brightness disk light than inner concentrated 
light. (5 systems)

 \item Edge-on disks: disk systems seen edge-on, and whose face-on morphology 
cannot be determined, but is presumably an S0 or spiral. (32 systems)
       
 \item Peculiar/irregular: Peculiars and irregulars are systems that 
appear to be disturbed, or peculiar looking, including elongated/tailed 
sources. These galaxies are possibly in some phase of a merger 
(Conselice et al. 2003a), or are dominated by star formation (502 systems).
   
 \item Unknown/too-faint: If a galaxy is too faint for any reliable 
classification it was placed in this category. Often these galaxies appear 
as smudges without any structure. These could be disks or ellipticals, but 
their extreme faintness precludes a reliable classification. (77 systems)
\end{enumerate}

\subsection{Extended CAS Structural Analysis}

\begin{figure}
 \vbox to 120mm{
\includegraphics[angle=0, width=90mm]{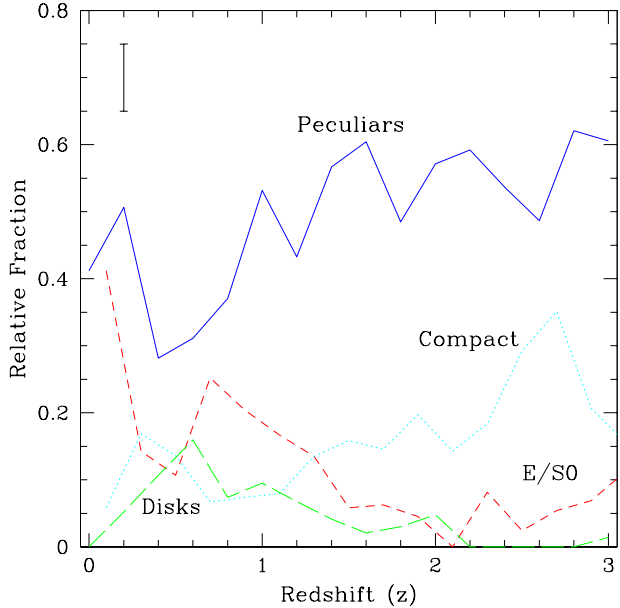}
 \caption{The relative distribution of galaxy types in the Hubble
Ultra Deep Field for systems selected with $z_{850} < 27$. Labelled
are disks, ellipticals/S0s, compact galaxies, and peculiars. Note
that the compact galaxies become an  important population at higher
redshifts.}
} \label{sample-figure}
\end{figure}

We use the CAS (concentration, asymmetry, clumpiness) parameters to probe the 
structures of our galaxies quantitatively.  The CAS parameters are a 
non-parametric method for measuring the forms of galaxies on resolved 
CCD images (e.g., Conselice et al. 2000a; Bershady et al. 2000; 
Conselice et al. 2002;  Conselice 2003).   The
basic idea is that galaxies have light distributions
that reveal their past and present formation modes (Conselice 2003). 
Furthermore, well-known galaxy types in the nearby universe fall in 
well defined regions of the CAS parameter space.  For example, the 
selection $A > 0.35$ locates systems which are nearly all major
galaxy mergers in the nearby universe (e.g., Conselice et al. 2000b; 
Conselice 2003; Hernandez-Toledo et al. 2005; Conselice 2006b).
In addition to the classic CAS parameters, we also investigate
the use of the similar Gini and M$_{20}$ parameters (Lotz et al. 2006) for
understanding the morphologies of the UDF galaxies.  We give
a brief description of these parameters below.

The way we measure structural parameters on the UDF image varies slightly
from what has been done earlier in the Hubble Deep Field, and GOODS 
imaging (e.g., Conselice et al. 2003a; Conselice et al. 2004). The basic
procedure, after cutting out the galaxy into a smaller image, is to
first measure the radius in which the parameters are computed. The
radius we use for all our indices is the Petrosian radii, which is the
radius defined as the location where the surface brightness at a given
radius is 20\% of the surface brightness within that radius (e.g.,
Bershady et al. 2000; Conselice 2003).   We use circular apertures for
our Petrosian radii and quantitative parameter estimation.  We begin
our estimates of the galaxy centre for the radius measurement at
the centroid of the galaxy's light distribution.  Through modelling
and various tests, it can be shown that the resulting radii do not
depending critically on the exact centre, although the CAS and other
parameters do (Conselice et al. 2000; Lotz et al. 2004).  The exact
Petrosian radius we use to measure our parameters is

$$R_{\rm Petr} = 1.5 \times r(\eta = 0.2),$$

\noindent where $r(\eta = 0.2)$ is the radius where the surface
brightness is 20\% of the surface brightness within that radius.

A very important issue, especially for the faint galaxies seen in the
UDF, is how to account for background light and noise. For faint galaxies there
is a considerable amount of noise added due to the sky, which must
be corrected.  Through various test, outlined in detail
in Conselice et al. (2008, in prep), we conclude that the proper
way to correct parameters for the background requires that the selected
background area be close to the object of interest. This is only an issue
for faint galaxies, and for galaxies imaged on large mosaics which
have a non-uniform weight map, and whose noise characteristics vary
across the field.  By using a background near each object we alleviate these
issues as the noise properties do not vary significantly over 
$\sim 0.5 - 1$ arcmin, where the galaxy and the
background area are selected.  We review below how the CAS and Gini/\m20
parameters are measured. For more detail see Bershady et al. (2000),
Conselice et al. (2000), Conselice (2003) and Lotz et al. (2006).

\subsubsection{Asymmetry}

The asymmetry of a galaxy is measured by taking an original galaxy 
image and rotating it 180 degrees about the galaxy centre, and then
subtracting the two images (Conselice 1997). There are corrections done for
background, and radius (explained in detail in Conselice et al. 2000a).
Most importantly, the centre for rotation is decided by an iterative
process which finds the location of the minimum asymmetry.  The formula
for calculating the asymmetry is given by:

\begin{equation}
A = {\rm min} \left(\frac{\Sigma|I_{0}-I_{180}|}{\Sigma|I_{0}|}\right) - {\rm min} \left(\frac{\Sigma|B_{0}-B_{180}|}{\Sigma|I_{0}|}\right)
\end{equation}

\noindent Where $I_{0}$ is the original image pixels, $I_{180}$ is the image
after rotating by 180\deg.  The background subtraction using light from a
blank sky area, called $B_{0}$, are critical for this process, and must 
be minimised in the same way as the original galaxy itself.  A lower value 
of $A$ means that a galaxy has a 
higher degree of rotational symmetry which tends to be found in
elliptical galaxies.
Higher values of $A$ indicate an asymmetric light distribution, which are 
usually found in spiral galaxies,  or in the more extreme case, merger 
candidates. The upper and lower bound for $A$, in this study are 1.18 
and $\sim 0$. The mean $A$ value is 0.26.

\subsubsection{Concentration}

Concentration is a measure of the intensity of light contained within a 
central region in comparison to a larger region in the outer-parts of a
galaxy.  The exact definition is the ratio of two circular radii which 
contain 20\% and 80\% ($r_{20}$, $r_{80}$) of the total galaxy flux,

\begin{equation}
C = 5 \times {\rm log} \left(\frac{r_{80}}{r_{20}}\right).
\end{equation}

\noindent This index is sometimes called C$_{28}$.  
A higher value of $C$ indicates that a larger amount of light 
in a galaxy is contained within a central region.  The upper and lower 
bounds of $C$ for 
our galaxy sample are $C = 1.8 - 4.4$, with a mean of $C = 2.77$.  This
particular measurement of the concentration correlates well with 
the mass and halo properties of galaxies (e.g., Bershady et al. 2000; 
Conselice 2003; Courteau et al. 2007).

\begin{figure}
 \vbox to 120mm{
\includegraphics[angle=0, width=90mm]{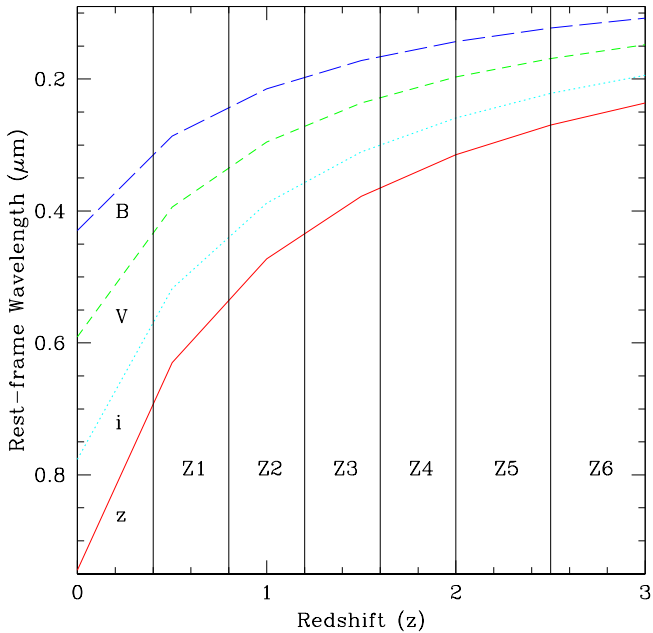}
 \caption{The rest-frame wavelength probed as a function of redshift
for each of our filters used in this study - $B_{450}$, $V_{606}$,
$i_{775}$ and $z_{850}$.  The vertical line denote the redshift
ranges we use to divide our sample into various redshift
cuts.}
} \label{sample-figure}
\end{figure}

\subsubsection{Clumpiness}

The clumpiness (sometimes called smoothness) $S$ is a parameter used to 
describe 
the fraction of light in a galaxy which is contained in clumpy light
concentrations.   Clumpy galaxies have a relatively large amount of
light at high spatial frequencies, 
whereas smooth systems, such as elliptical galaxies contain light at low 
spatial frequencies. Galaxies which are undergoing star formation tend to 
have very clumpy structures, and high $S$ values.  Clumpiness can be 
measured in a number of ways, the most common method used,
as described in Conselice (2003) is,

\begin{equation}
S = 10 \times \left[\left(\frac{\Sigma (I_{x,y}-I^{\sigma}_{x,y})}{\Sigma I_{x,y} }\right) - \left(\frac{\Sigma (B_{x,y}-B^{\sigma}_{x,y})}{\Sigma I_{x,y}}\right) \right],
\end{equation}

\noindent where, the original image $I_{x,y}$ is blurred to produce 
a secondary image,  $I^{\sigma}_{x,y}$.  This blurred image is
then subtracted from the original image leaving a 
residual map, containing only high frequency structures in
the galaxy (Conselice 2003). To quantify this, we normalise the
summation of these residuals by the original galaxy's total light, and
subtract from this the residual amount of sky after smoothing
and subtracting it in the same way.  The size of the smoothing kernel 
$\sigma$ is
determined by the radius of the galaxy, and is $\sigma = 0.2 \cdot 1.5
\times r(\eta = 0.2)$ (Conselice 2003).  Note that the centres of galaxies are
removed when this procedure is carried out.  

\subsubsection{Gini Coefficient}

The Gini coefficient is a statistical tool originally used in economics to 
determine the distribution of wealth within a population, with higher values 
indicating a very unequal distribution (Gini of 1 would mean all wealth/light 
is in one person/pixel), wile a lower value indicates it is distributed 
more evenly amongst the population (Gini of 0 would mean everyone/every pixel 
has an equal share).   The value of G is defined by the Lorentz curve 
of the galaxy's light distribution, which does not take into 
consideration spatial position.  Each pixel is ordered by its
brightness and counted as part of the cumulative distribution (see
Lotz et al. 2004, 2006).    The mean value of Gini in our UDF catalogue is 
0.71.

\subsubsection{M$_{20}$}

The \m20 parameter is a similar parameter to the concentration in that it 
gives a value that indicates 
whether light is concentrated within an image; it is however calculated 
slightly differently.  The total moment of light is calculated by summing the 
flux of each pixel multiplied by the square of its distance from the centre.  
The centre is deemed to be the location where \m20 is minimised
(Lotz et al 2004).  The value of \m20 is the moment of the 
fluxes of the brightest 20\% of light in a galaxy, which is
then normalised by the total light moment for all pixels (Lotz
et al. 2004, 2006)

\begin{figure*}
 \vbox to 150mm{
\includegraphics[angle=0, width=184mm]{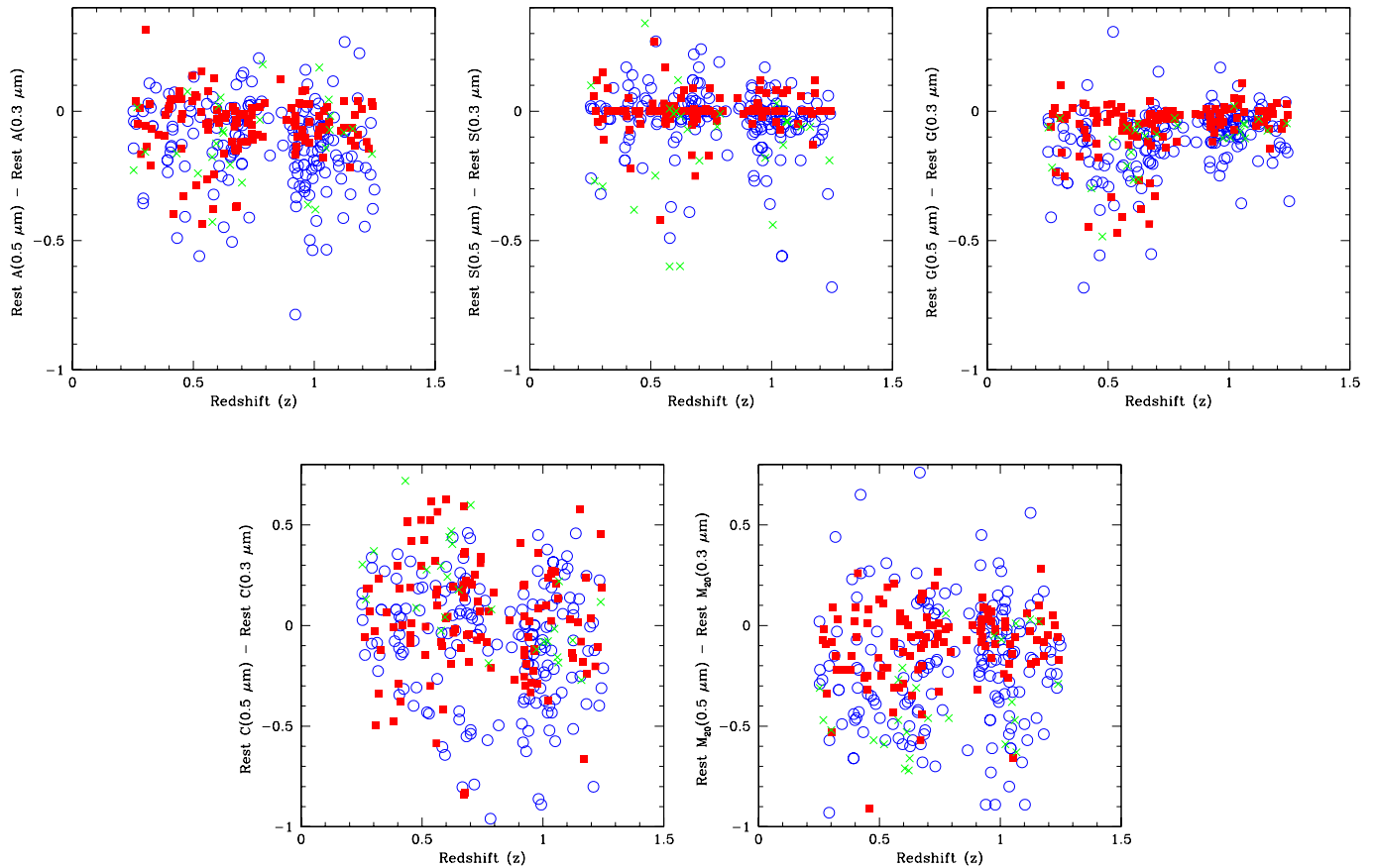}
 \caption{The change in the CAS and Gini and \m20 parameters
as a function of redshift. We trace in this figure the change
between the rest-frame 0.5 $\mu$m and 0.3 $\mu$m morphology.  The 
concentration and
\m20 values have a much larger dynamical range than the CAS and Gini
values which is intrinsic to the way these are measured (see text). In 
relative 
terms these changes are similar within the other parameters. The symbols are
the same as in Figure~1.}
} \label{sample-figure}
\end{figure*}
\vspace{4cm}

The main differences between \m20 and $C$ are due to the moments in
\m20 which depend on the distance from the galaxy centre. The value of
\m20 will therefore be more affected by spatial variations, and also the 
centre of the galaxy is again a free parameter.  This can make it more 
sensitive to possible mergers. In our study we find that the upper and 
lower bounds of \m20 are \m20 $= 0$ to \m20 $= -2.51$, with a mean of 
\m20 $=-1.45$.  

\subsection{Overview of Measured CAS/G/\m20 Values}

We apply a revised CAS system to our UDF galaxy sample
to determine their structural parameters.  There are several caveats to 
using the ACS imaging to measure these parameters.  The first is that 
there are 
morphological $k-$correction and surface brightness dimming effects which will 
change the measured parameters, such that the 
asymmetry and clumpiness indices will decrease (Conselice et al. 2000a; 
Conselice 2003), and the concentration index will be less reliable (Conselice
2003).  There is also the issue that for systems at $z > 1.2$ we
are viewing these galaxies in their rest-frame ultraviolet using ACS
data, which
means that there are complications when comparing their measured structures
with the calibrated rest-frame optical indices for nearby galaxies.  We
deal with the first problem in \S 4.2 and discuss the reliability of
the measurements themselves below.

To test the reliability of our parameters, we plot all five indices
against apparent magnitude to our $z_{850} = 27$ magnitude limit.  
We find that there is no dependence on 
magnitude for the morphological parameters within this limit. 
At magnitudes fainter than our limit we find that all of our structural 
parameters become systematically smaller or higher, simply just due to a 
lower signal
to noise. Our limit of $z_{850} < 27$ ensures that we are within the
regime where systematic problems are not dominating our signal.

\vspace{1cm}
\setcounter{table}{0}
\begin{table}
 \caption{The change in CAS and Gini/\m20 parameters from the rest-frame
optical to near-UV as function of redshift.  These values are defined
such that $\Delta = \lambda(0.5 \mu m) - \lambda (0.3 \mu m)$}
 \label{tab1}
 \begin{tabular}{@{}lrr}
  \hline
\hline
Peculiars & $z = 0.25 - 0.75$ & $z = 0.75 - 1.25$ \\
\hline
$\frac{\Delta C}{\Delta \lambda} (\mu m^{-1})$ & -0.09$\pm$1.50 & -0.78$\pm$2.30 \\
$\frac{\Delta A}{\Delta \lambda} (\mu m^{-1})$ & -0.26$\pm$0.91 & -0.83$\pm$1.06 \\
$\frac{\Delta S}{\Delta \lambda} (\mu m^{-1})$ & -0.48$\pm$0.57 & -0.33$\pm$0.61 \\
$\frac{\Delta G}{\Delta \lambda} (\mu m^{-1})$ & -0.65$\pm$0.65 & -0.39$\pm$1.22 \\
$\frac{\Delta M_{20}}{\Delta \lambda} (\mu m^{-1})$ & -0.91$\pm$1.50 & -1.00$\pm$1.94 \\
\hline
Elliptials & $z = 0.25 - 0.75$ & $z = 0.75 - 1.25$ \\
\hline
$\frac{\Delta C}{\Delta \lambda} (\mu m^{-1})$ & 0.43$\pm$1.50 & -0.11$\pm$1.50 \\
$\frac{\Delta A}{\Delta \lambda} (\mu m^{-1})$ & -0.26$\pm$0.61 & -0.22$\pm$0.56 \\
$\frac{\Delta S}{\Delta \lambda} (\mu m^{-1})$ & -0.04$\pm$0.43 &  0.06$\pm$0.33 \\
$\frac{\Delta G}{\Delta \lambda} (\mu m^{-1})$ & -0.43$\pm$0.78 & -0.11$\pm$0.61 \\
$\frac{\Delta M_{20}}{\Delta \lambda} (\mu m^{-1})$ & -0.69$\pm$1.26 & -0.21$\pm$0.69 \\
\hline
Spirals & $z = 0.25 - 0.75$ & $z = 0.75 - 1.25$ \\
\hline
$\frac{\Delta C}{\Delta \lambda} (\mu m^{-1})$ & 1.30$\pm$0.87 & -0.33$\pm$2.30 \\
$\frac{\Delta A}{\Delta \lambda} (\mu m^{-1})$ & -0.57$\pm$0.57 & -0.44$\pm$1.40 \\
$\frac{\Delta S}{\Delta \lambda} (\mu m^{-1})$ & -0.61$\pm$1.13 & -0.56$\pm$1.20 \\
$\frac{\Delta G}{\Delta \lambda} (\mu m^{-1})$ & -0.70$\pm$0.57 & -0.33$\pm$0.83 \\
$\frac{\Delta M_{20}}{\Delta \lambda} (\mu m^{-1})$ & -2.30$\pm$1.00 & -1.04$\pm$1.70 \\
\hline

 \end{tabular}
\end{table}

\subsection{Stellar Masses}

The stellar masses we measure are computed using the techniques
described in Bundy et al. (2006) and Conselice et al. (2007b) using the
BV$iz$JH data.  The basic method we use consists of fitting a grid of model
SEDs constructed from Bruzual \& Charlot (2003) (BC03) stellar population 
synthesis models, with different star formation histories.  We use an 
exponentially declining model to characterise the star formation, 
with various ages, metallicities and dust contents included.  
These models are parameterised by an age, and an e-folding time for 
parameterising the  history of star formation.  These parameterisations are
fairly simple, and it remains possible that stellar mass from
older stars is missed under brighter younger populations.   However, 
stellar masses measured through our technique are roughly the expected 
factor of 5-10 smaller than dynamical masses at $z \sim 1$ using a sample 
of disk galaxies (Conselice et al. 2005b), demonstrating their inherent 
reliability.  We furthermore test how these stellar masses would
change utilising the newer Bruzual \& Charlot (2007) models, finding
at most a 0.07 dex decrease due to the newer implementation of
thermal-pulsating AGB stars (see also Conselice et al. 2007b).

We calculate  the likely stellar mass, age, and absolute magnitudes for 
each galaxy at all star formation histories, and determine stellar
masses based on this distribution.  Distributions with larger ranges of 
stellar masses have larger resulting uncertainties. It turns out that 
while parameters such as the age, e-folding time, metallicity, etc. are not 
likely accurately measured in these calculations, due to various 
degeneracies, the stellar mass is robust.  Typical errors for our 
stellar masses are 0.2 dex from the width of the probability distributions.  
There are also uncertainties from the choice of the IMF.  Our stellar masses 
utilise the Salpeter IMF, which can be converted to Chabrier IMF stellar 
masses by subtracting 0.25 dex.  There are additional
random uncertainties due to photometric errors.  The resulting
stellar masses thus have a total random error of 0.2-0.3 dex,
roughly a factor of two. This fitting method gives similar results
used to compute merger fractions as a function of stellar mass in
previous work (Conselice et al. 2003a). 

\section{Results}

\subsection{General Features}

By plotting our visual estimates of morphology vs. various
properties, we can decipher the formation modes of galaxies found at a 
faint $z_{850}$ magnitude limit.  Figure~1 plots stellar mass versus
redshift for our sample, with the various morphological types labelled. 
All galaxy types are seen at all redshifts, and at
nearly all masses.  As can
also been seen, there is a slight increase in the upper envelope
of stellar masses at $z < 1$. While we find galaxies with
masses M$_{*} >$ \lmass at all redshifts, in general there are
few galaxies with masses M$_{*} >$ \mass at any redshift, 
demonstrating how difficult it is to study these systems within
small fields like the UDF and HDFs (Conselice et al. 2007b).

\begin{figure*}
 \vbox to 150mm{
\includegraphics[angle=0, width=184mm]{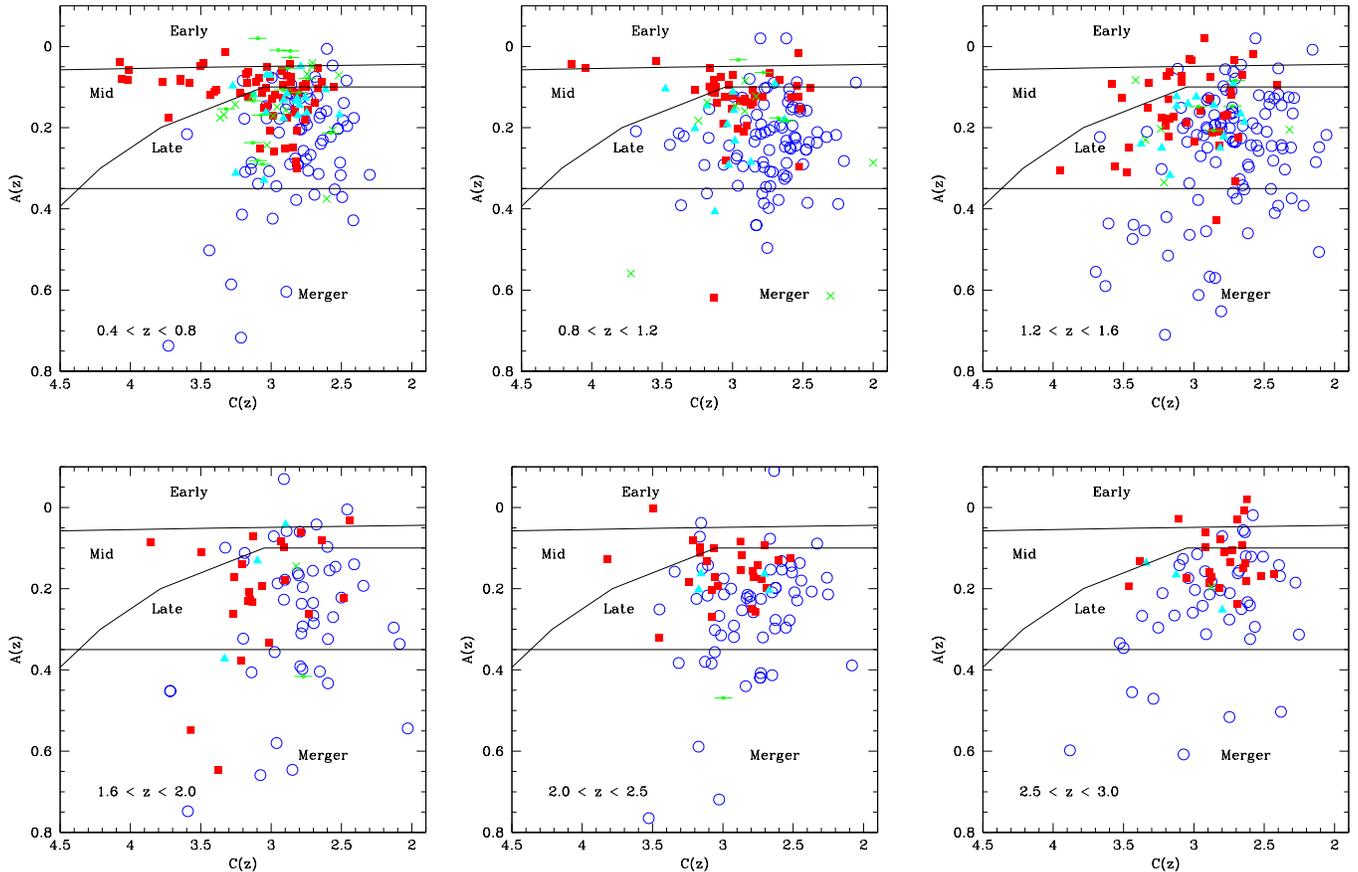}
 \caption{The distribution of our galaxy sample in the UDF as
seen in the concentration-asymmetry plane.  The lines on each plot
denote the region in which different galaxy types, as seen in the
rest-frame optical are found in the nearby universe (e.g., Bershady et 
al. 2000; Conselice 2003).  The redshift range for each panel follows
the division seen in Figure~3, and is labelled at the bottom of each
panel.  The symbol types for the points are the same as in Figure~1.}
} \label{sample-figure}
\end{figure*}
\vspace{7cm}

We also plot
the fraction of types at our $z_{850}$ limit as a function of
redshift in Figure~2. Figure~2 shows that galaxies that look peculiar by eye
dominate the galaxy population at all redshifts within a $z_{850} = 27$
magnitude selection limit. This is especially
true at higher redshifts, where galaxies that look unusual make
up roughly half of the galaxy population.
This appears to be a different morphological distribution with redshift
than what has been seen in the past using absolute magnitudes and
stellar masses (e.g., Conselice et al. 2005a). We can explain this difference 
through our use of an apparent
magnitude limit, rather than an absolute magnitude limit.  
It has been know since the first deep HST imaging (e.g., Driver
et al. 1998; Glazebrook et al. 1995) that at fainter magnitudes
there are more peculiar galaxies.  There are also more fainter
galaxies than brighter ones, thus it is not surprising that peculiars
are the dominant population (e.g., Elmegreen et al. 2005; Ravindranath et al. 2006).    
If we do a stellar mass cut at a high
limit, such as M$_{*} > 10^{9.5}$ \solm, we find that the peculiars 
no longer dominate the population.

The other reason for the apparently large number of peculiars is that
the higher resolution of the ACS camera makes it easy to detect
peculiar features that otherwise would not be identifiable within 
WFPC2 imaging.  This is particularly true for galaxies that would
be considered spirals or disks, but lack a coherent obvious
structure in their morphological appearance as seen within ACS.

It is also worth noting a few more interesting features of Figure~2.
The first is that while the peculiars dominate the galaxy population
within the $z_{850} < 27$ selection, the ellipticals and disks
decline in their relative contributions at lower redshifts.  However,
at redshifts $> 1.5$ the galaxies classified
by eye as compacts are an important part of the galaxy population,
with a fraction of 20-35\%.  This is a significant fraction of
the galaxy population, particularly at $z > 2$ and it is worth
briefly describing what these systems are.  From Figure~1 we can
see that these galaxies have high stellar masses
with M$_{*} >$ \llmass, and often M$_{*} >$ \lmass.  These systems could
be the progenitors of ellipticals seen at lower-redshifts,
or could be a separate galaxy population that is in some form of evolution.

\subsection{The Morphological K-Correction}

One of the major problems with studying galaxy structure and
morphology is that these features can depend quite strongly
on the rest-frame wavelength probed (e.g., Hibbard \& Vacca  1997;
Windhorst et al. 2002;
Papovich et al. 2003; Taylor-Mager et al. 2007).  To understand
this issue within the UDF, we determine how our quantitative indices 
change as a function of wavelength at $z < 1$, where we probe 
light from the rest-frame ultraviolet to the rest-frame B-band 
for the same galaxies, using our four ACS filters.

For various reasons it is important to determine how galaxy 
structure changes as a function of
wavelength, and the UDF, due to its depth, is the ideal place to carry
this out by directly probing these differences.  We can furthermore 
use these results later when we do not
have a full set of ACS-UDF filters, or as often happens, we do not
have a filter which probes the rest-frame optical at higher redshifts. 
Until the advent of a high resolution near-infrared camera, such as WFC3 on
Hubble, ACS optical imaging is our only probe of high resolution 
imaging at $z > 1.2$.    

We calculate morphological K-corrections in a number of ways.
The first way is to simply determine how the CAS, Gini, and
\m20 parameters change as a function of wavelength and redshift.
We define a quantify, $\Delta P_{\lambda}$, which is the change in the
quantitative parameter
values per unit wavelength ($\mu$m). The values of $\Delta P_{\lambda}$ 
change as a function of rest-frame wavelength ($\lambda$), redshift ($z$), 
quantitative parameter (CAS,G,\m20), and the eye-ball estimates of the 
morphological type ($T$).  We quantify the morphological K-correction by,

\begin{equation}
\Delta P_{\lambda} (\lambda, z, T) = \frac{\delta P(\lambda_{1} - \lambda_{2})}{\delta \lambda},
\end{equation}

\begin{figure*}
 \vbox to 150mm{
\includegraphics[angle=0, width=184mm]{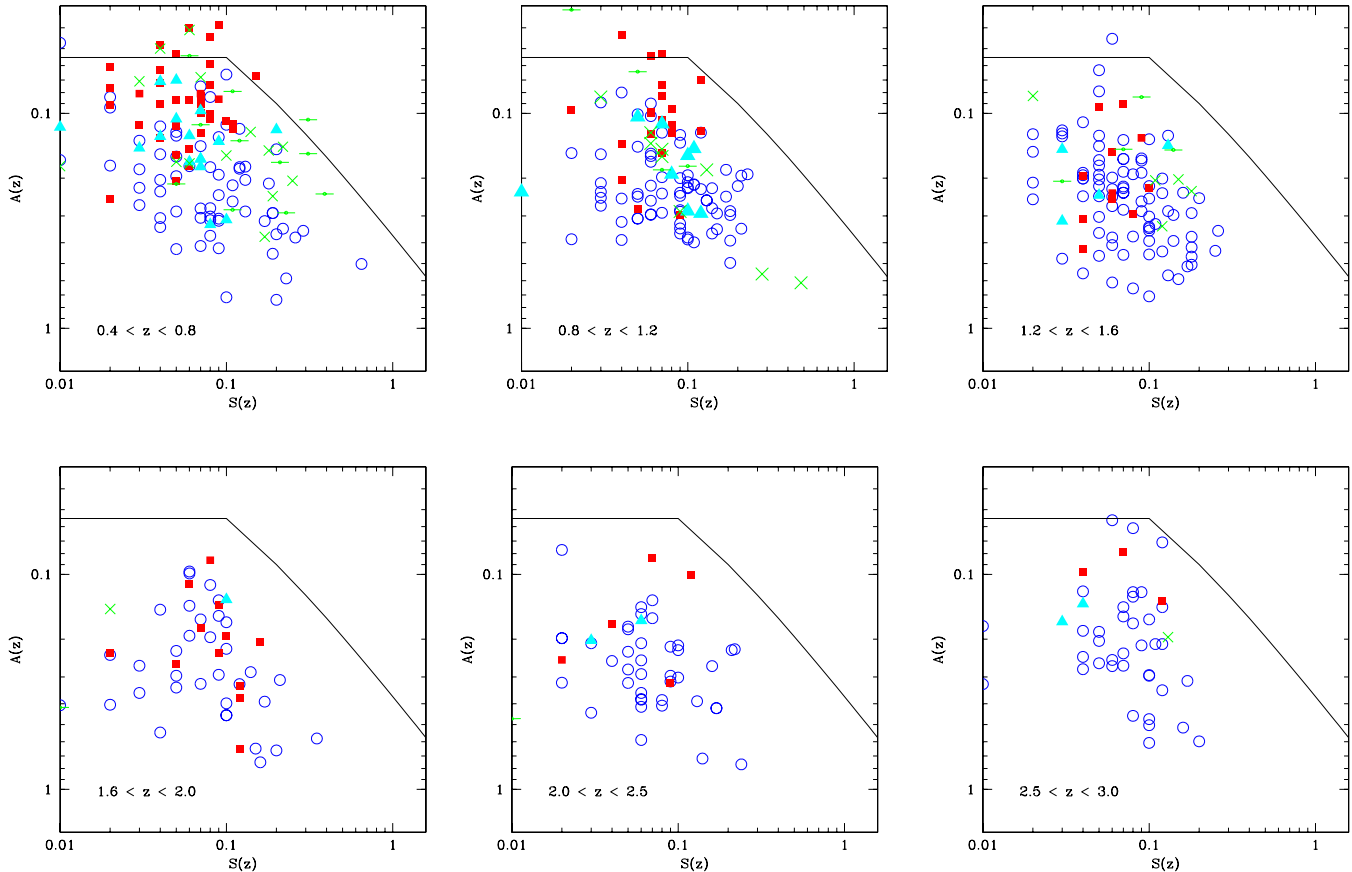}
 \caption{The relation between the asymmetry ($A$) and clumpiness ($S$)
for our sample of UDF galaxies in the observed $z_{850}$ band.    
The solid line shows the relationship
between these two parameters as found for nearby galaxies that are not
involved in mergers (Conselice 2003). This relationship is such that
non-merging galaxies with a higher clumpiness has a slightly higher
asymmetry, both due to star formation. Merging galaxies, where 
the structure is distorted due to bulk asymmetries from a merger
have a larger asymmetry for their clumpiness.  The point symbols
are the same as in Figure~1 and the redshift range for each individual
panel is listed. }
} \label{sample-figure}
\end{figure*}
\vspace{4cm}

\noindent where $\delta P$ is the change in the parameter of interest between
$\lambda_{1}$ and $\lambda_{2}$, and $\delta \lambda$ is the
change in rest-frame wavelength measure in microns.  Before describing
how $\Delta P_{\lambda}$ varies with redshift and quantitative parameter, 
it is important to re-motivate why we are interested in this quantity.

\begin{figure*}
 \vbox to 150mm{
\includegraphics[angle=0, width=184mm]{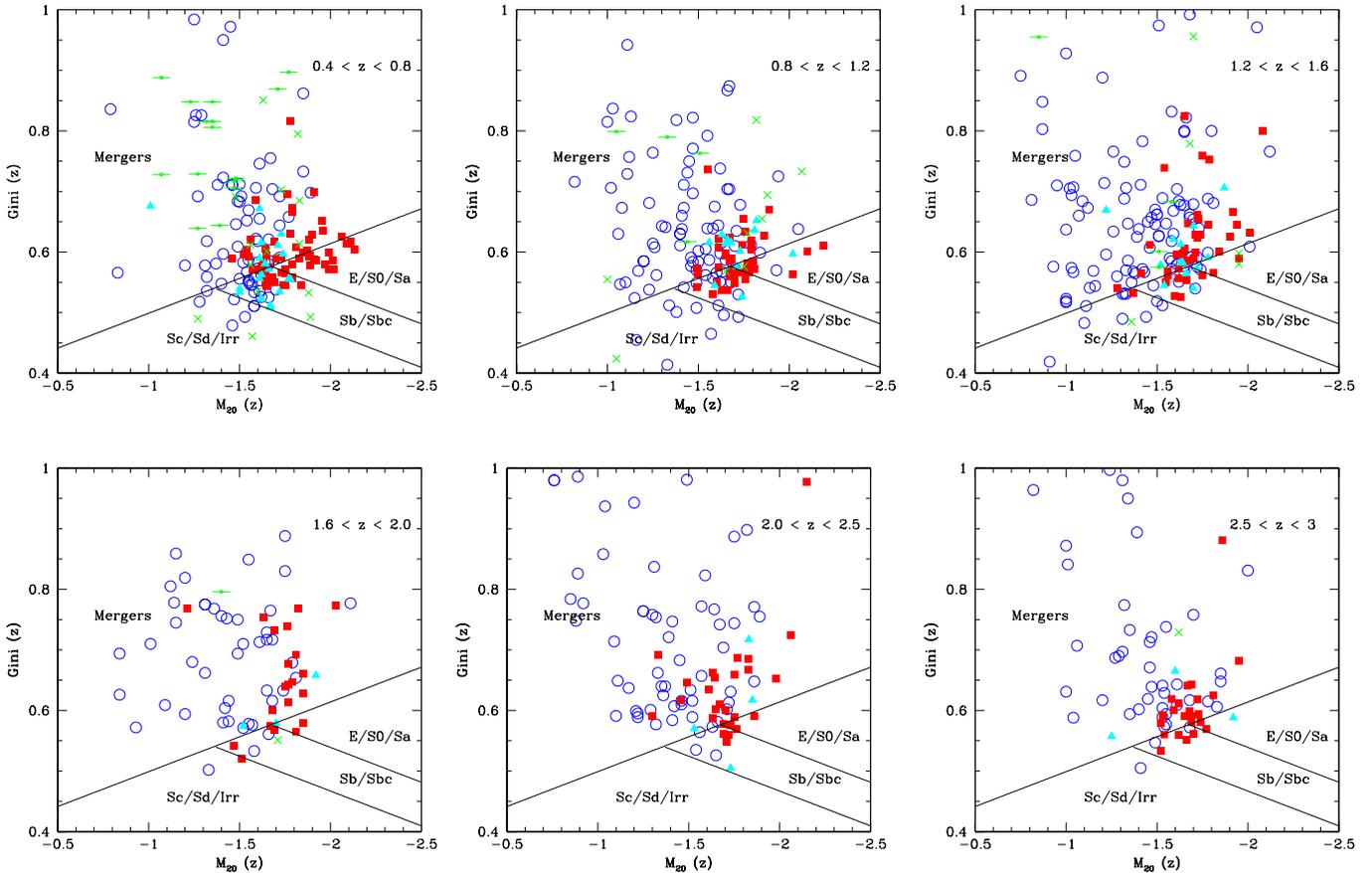}
 \caption{The relationship between the Gini and \m20 index in the
observed $z_{850}$ band for our sample of UDF galaxies.  The lines and
symbols denote the region where different nearby galaxy types are
found (Lotz et al. 2006).  The redshift range for each individual panel
are the same as in Figure~5 \& 6. The symbols for the points are the
same as in Figure~1 }
} \label{sample-figure}
\end{figure*}

\vspace{4cm}

The highest redshifts we probe in this paper are $z \sim 2.5-3$.
When we observe galaxies at these redshifts in our reddest observed
band ($z_{850}$), we are viewing them in the rest-frame ultraviolet at
$\sim$ 2700 \AA\, (see Figure~3).  Due to our suite of four ACS filters, 
this same rest-frame wavelength is probed by the observed $B_{450}$-band for 
galaxies 
at $z \sim 0.5$, the $V_{606}$-band for galaxies at $z \sim 1$, and the 
$i_{775}$-band for galaxies at $z \sim 2$ (Figure~3).    
We can therefore determine at lower redshifts how quantitative
indices may change between the rest-frame optical and rest-frame 
ultraviolet at higher redshifts as a function of apparent morphological
type (see Conselice et al. 2005a for 
a direct measurement of this using NICMOS imaging in the Hubble
Deep Field).

Figure~4 and Table~1 show the results of this investigation.  We 
plot and list the differences between the values of our
parameters in the rest-frame 0.3 $\mu$m, and at
rest-frame 0.5 $\mu$m (Figure~4).  We limit this analysis to
those galaxies which are at $z < 1.25$, as at higher redshifts
we are no longer probing above the 4000 \AA\ break (Figure~3).
For this analysis we take the values in the observed
$z_{850}$ and $V_{606}$ bands for galaxies between $z = 0.75 - 1.25$,
and $i_{775}$ and $B_{435}$ for galaxies between $z = 0.25 - 0.75$.
This lets us probe the quantitative structural $k-$correction
between the rest-frame near-ultraviolet at $< 300$ nm, and the
rest-frame B-band at $\lambda \sim 500$ nm.

We find  that the change in
galaxy quantitative parameters between the rest-frame UV and the
rest-frame optical, $\Delta P_{\lambda}$, varies amongst morphological type,
as well as redshift.  As Figure~4 shows, the galaxy type with the
most variation between optical and near-UV are the peculiars.  For
the asymmetry and clumpiness indices we find that the value of 
$\Delta P_{\lambda}$ is largest for the peculiar galaxies at 
higher redshifts, compared with other morphological types.   We do
not find such a large differential between galaxy types within the
other parameters. This reveals that this difference between the
$A$ and $S$ parameters for the peculiars is likely driven at least
partially by star formation present within these systems.

Note that the concentration and \m20 values have a larger dynamical
range than the $A$, $S$ and Gini parameters, which accounts for
the larger differences seen (Figure~4).  However, when we normalise these
differences by their initial values we find that the range
is relatively similar to the other parameters, particularly
for the concentration index, whose values only change by
$\delta C/C = -0.1$ on average.

As Table~1 shows, most of the calculated values of $\Delta P_{\lambda}$ are similar
at $z = 0.25 - 0.75$ as at $z = 0.75 - 1.25$.  The exceptions to this
are the concentration, \m20 and Gini indices for the spirals, and the 
concentration index for both the peculiars and ellipticals, and the asymmetries
for the peculiars. Perhaps surprisingly, the most stable parameter is
the $S$ index, which has the smallest value of $\Delta P_{\lambda}$ of
all the parameters.  This is likely due to the fact that the high
resolution and high depth of the ACS images allows us to reveal the
fine structure in these galaxies at all observed wavelengths.

The likely reason for these changes, at least in terms of the concentration
index, is that the star formation rate for these galaxies, and all
galaxies at $z < 1$, is declining. This results in the differences between
a blue and a UV band becoming even more pronounced, as there is not as much 
star formation to create a large signal in the ultraviolet. This is the
reason the ellipticals change in concentration from a negative
gradient to a relatively large positive one.  Since star formation
is seen in ellipticals at $z \sim 1$ (e.g., Stanford et al. 2004; 
Teplitz et al. 2006), we
are likely witnessing a lack of star formation at the lower redshifts.
We utilise these results in the later discussion of this paper on 
the merger fraction for these galaxies, and how to interpret
structural indices at higher redshifts.

\subsection{Quantitative Structure}

There are a few very popular method for `classifying' galaxies
via quantitative parameters. These are the concentration-asymmetry
plane (e.g., Bershady et al. 2000), the asymmetry-clumpiness
plane (Conselice 2003), and the Gini/\m20 plane (Lotz et al.
2004, 2006).  We investigate where eye-ball estimates of
galaxy types fall into these different areas of quantitative 
structural space at $0.4 < z < 3$,
as well as how the various values agree, or otherwise,
with each other.

\subsubsection{The Concentration-Asymmetry plane}

Figure~5 shows the UDF concentration-asymmetry plane, without redshift 
corrections applied (Conselice 2003), as observed in the $z_{850}$-band.
The lines on this diagram denote the general area where early/mid/late and 
merging galaxies are located in this space (Bershady et al. 2000; Conselice
2003). One of the most striking aspects of Figure~5 is that very few galaxies 
appear in the early region; this is due to a  higher general level of 
asymmetry within UDF galaxies in comparison to the local universe.  
Interestingly, this is true of the galaxies classified as early types, 
a very small percentage of which actually fall within the early region at
any redshift. 

The disk galaxies, as classified by eye, appear in the late area of the plot 
where they are  
expected to be found, although there are a few that appear above the merger 
limit of $A = 0.35$.  These disk galaxies are always very clumpy systems which
appear to have multiple components.  The edge-on galaxies are 
however distributed quite randomly around the plot, but they generally contain
a low asymmetry.   Most are located in the late-type region on the
CA space.

Another interesting trend is that the disturbed ellipticals (the cyan triangles
in Figure~5)
fall in a  buffer zone between early-types and the peculiar galaxies. Most
of them are in fact within the late-type region, although these galaxies
do not have a disk morphology.  This  suggests that
these galaxies are within an intermediate state of formation between the
peculiar galaxies and the relaxed ellipticals.  Figure~5 also shows
how the concentration-asymmetry plane varies with redshift.  The concentrated 
early-types do not appear to exist at higher redshifts. While at 
$0.4 < z < 0.8$ 
there are many early-types with a high light concentration, very few of these 
systems are found at higher redshift.   This is likely an effect
of their formation, which can also be seen in the growth of stellar
masses for early-types, where few very massive systems are
found at $z > 1$ (Figure~1). 

As Figure~5 shows,  the asymmetry index is very good at finding galaxies
which are classified by eye as peculiar/merger.  There are only a few cases
where galaxies which were classified as something other than a peculiar
are found in the region where mergers are expected, $A > 0.35$.  In summary
at $0.4 < z < 3$ we find that 86$\pm10$\% of galaxies which are at $A > 0.35$ 
are classified as peculiars.   This method also finds 20$\pm3$\% of the 
galaxies classified as peculiars are within 
the $A > 0.35$ region, with 82 of the classified peculiars out of 410 in the
$A > 0.35$ regime.  These relative fractions do not change significantly
at different redshifts up to $z \sim 3$.

When we go to higher redshifts a few interesting patterns are
seen. The first is that the classified ellipticals, on average, have
a higher asymmetry, and lower concentration.  Within $0.4 < z < 0.8$
the average values are $<A> = 0.13$ and $<C> = 3.1$.  Up to $z \sim 1.6$
these values remain roughly the same.  However at $1.6 < z < 2$
these values change to $<A> = 0.21$ and $<C> = 3.1$.  While the
average concentration remains the same, the average asymmetry is
much higher. As we are probing the near-UV with this filter at these
redshifts, this likely shows that these systems are becoming more
dominated by star formation at this epoch.  

\begin{figure}
 \vbox to 120mm{
\includegraphics[angle=0, width=90mm]{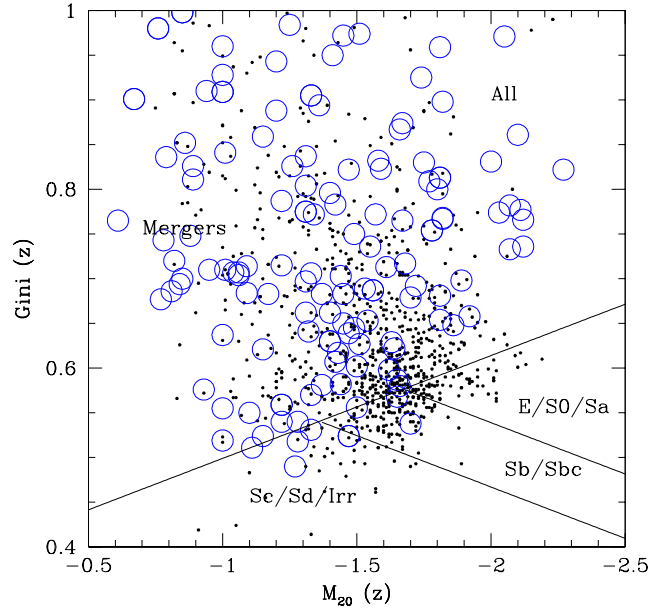}
 \caption{The Gini vs. \m20 diagram for our entire UDF sample
at all redshifts in the observed $z_{850}$ band.  The solid blue 
circles show the location of galaxies which are identified through
the CAS method as a major merger, $A > 0.35$.  The small points show
those objects which have an asymmetry value of $A < 0.35$.  As can
be seen, in general the galaxies identifiable as a merger within the
CAS system would also be identified as merger within the Gini-\m20
plane. However, there are objects within the merger region which
would not be identified as a merger within CAS.}
} \label{sample-figure}
\end{figure}

In general, we find a very similar pattern within the CA
space where the peculiars and early-types are found. While we
cannot know from this study whether these early-types are in fact
the progenitors of the early-types we see today, none the less the
fact that the visually classified early-types are systematically
less asymmetric than the peculiars, out to $z \sim 3$, shows that
the method of classification within the CA space generally provides
a good separation between galaxy types as identified by eye.

\subsubsection{Clumpiness-Asymmetry Plane}

Figure~6 shows the relation between clumpiness ($S$) and asymmetry ($A$)
for our sample of galaxies within the UDF.  The solid line shown is
the $z = 0$ relation between $S$ and $A$ for non-merging nearby
galaxies (Conselice 2003).  Generally, galaxies which fall along or near 
this line have
a structure which is dominated by star formation, which produces
high clumpiness values, as well as a higher asymmetry. As can
be seen in Figure~6, few of our galaxies appear to follow this line,
and most are more asymmetric at a given clumpiness than what is
expected based on the $z = 0$ relation.

There are two reasons why these galaxies appear to deviate from the
$z \sim 0$ relation between $A$ and $S$. The first is simply due
to resolution and S/N effects which will lower the level of 
asymmetry and clumpiness. Both the asymmetry and the clumpiness
will decline when galaxies are simulated at higher redshifts, however,
the clumpiness values are affected to a larger degree than the
asymmetry (Conselice 2003).   Simulations show that in general
the typical difference is $A - S \sim 0.2$.   Shifting galaxies
by 0.2 dex in $S$ however does not account for the fact that
so many galaxies do not fall near the $z \sim 0$ relationship
between these two quantities.

By examining the $0.4 < z < 0.8$ redshift range, we can get some
idea for which types of galaxies deviate most from the non-merging
galaxy relation, and perhaps understand why.  First, as can
be seen in Figure~6, most of the high-$S$ galaxies at $S > 0.1$
are either disk galaxies, or peculiars, while most of the
high-$A$ systems are peculiars.  When we apply  the
$A - S = 0.2$ redshift correction most of the disk
galaxies fall near, or within, the region of the $z \sim 0$ relation.
Not surprisingly, the galaxies which do not fall near this relation
are the peculiars, the peculiar ellipticals, and in some cases
the ellipticals themselves.  These trends generally remain, although
the peculiars and the ellipticals occupy a a very similar
part of parameter space at higher redshifts, which is
not the case for the $A-C$ plane where these types can
be distinguished (\S 4.3.1).

\begin{figure*}
 \vbox to 150mm{
\includegraphics[angle=0, width=184mm]{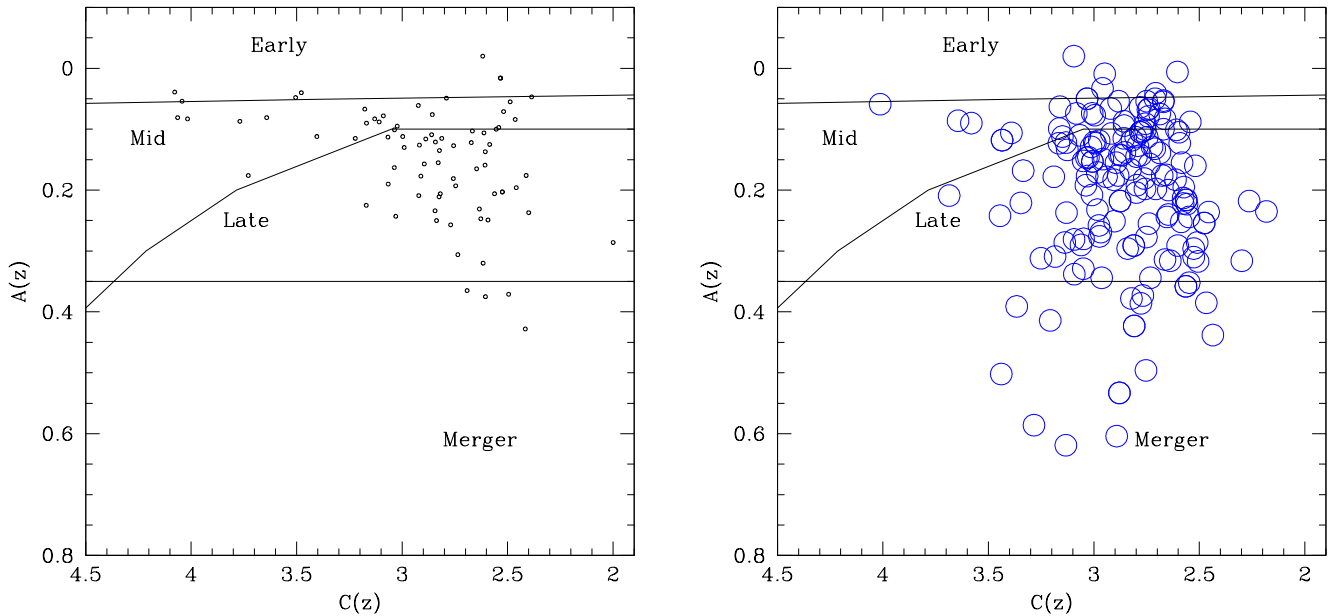}
 \caption{The concentration-asymmetry plane in the observed $z_{850}$
band at $z < 1$.  The left panel shows the location of galaxies which are 
identified as non-mergers within the Gini-\m20 plane.  The
right panel shows the location of galaxies in $CA$ space
that are identified as a merger within the Gini-\m20 plane.  As can
be seen, many of the Gini-\m20 mergers would be identified as
a non-merger within CAS space. }
} \label{sample-figure}
\vspace{-4cm}
\end{figure*}

\subsubsection{The Gini-\m20 plane}

In addition to the CAS values, the newer Gini and the \m20 coefficients 
also give information on morphological characteristics of galaxies (e.g.,
Abraham et al. 2003; Lotz et al. 2004; Lotz et al. 2006). We
show in Figure~7 the relation between Gini and \m20 for galaxies within
the UDF up to $z \sim 3$.  Areas to the right of this figure (in the
low \m20 regime) are where galaxies which are more centrally concentrated
in light are found.  For example, early-type galaxies appear 
on the right side of this space, as these galaxies have a very central light 
concentration.    

Similarly, galaxies with larger Gini indices indicate systems
where a large majority of the light is contained in few pixels.  While lower 
Gini values indicate galaxy images that have light profiles spread more 
evenly amongst all the pixels.  Using this frame-work, Lotz et al.
(2006) determined that various cuts in the Gini-\m20 space can be used
to separate galaxies of various types.  We plot these lines and the
galaxy types they denote on Figure~7.

The Gini-\m20 plane does a fairly good job of separating galaxy types
as identified by eye.  This is particularly so when examining galaxies
in the rest-frame optical at $z < 1.4$ (Figure~7). As can be
seen within the redshift range $0.4 < z < 0.8$, the peculiars generally
have a higher Gini index and are mostly found within the merger region
specified by Lotz et al. (2006).
However, as  shown in Figure~7 there are many galaxies classified
as a peculiar which are within the non-merger region, specifically in the
Sc/Sd/Irr and Sb/Sbc regime.  There are also many galaxies classified
as early-type and especially edge-on disk, which are within the
merger region.

 Within the $0.4 < z < 0.8$ redshift range
75$\pm10$\% of the galaxies classified by eye as peculiars are within the 
Gini/\m20 merger defined region.  However, only 44$\pm6$\% of the 
galaxies within the merger
regime at these redshifts are classified as peculiars.  Most 
other galaxy types are within this region, including nearly all of
the edge-on disks, ellipticals, and face-on disks themselves.  As
can be seen in Figure~7, the ellipticals/S0s/compacts are the only
galaxies found within the E/S0/Sa region of Gini-\m20 space.

At higher redshift, this remains the same up to $z \sim 2$. At
$0.8 < z < 1.2$ we find that 78$\pm10$\% of the peculiars are in the
mergers region, and 63$\pm8$\% of galaxies within the mergers region
are classified as peculiars.  The equivalent quantities at higher
redshifts are (fraction peculiar detected, fraction peculiar in region): 
$1.2 < z < 1.6$: (84$\pm9$\%, $66\pm7$\%); 
$1.6 < z < 2.0$: ($93\pm14$\%, $66\pm10$\%); $2.0 < z < 2.5$: 
($97\pm13$\%, $71\pm9$\%); $2.5 < z < 3.0$: ($93\pm15$\%, $67\pm10$\%).
Generally, the Gini-\m20 method identifies a higher fraction of
the peculiar galaxies as mergers, yet it has a higher contamination
of non-peculiar galaxies within the merger region. As has been known
from several investigations, it is necessary to clean a Gini/\m20
catalog of merger candidates before using it to determine the
merger fraction (Lotz et al. 2006), which is generally also the
case for the CAS method (De Propris et al. 2007).

What we find at higher redshifts is a similar pattern as at the
$0.4 < z < 0.8$ redshift range, although the early-types tend to
drift into the higher Gini and higher \m20 regime, revealing that
these galaxies are becoming less concentrated with time in the
observed $z_{850}$-band, as seen in the CAS parameters (\S 4.3.1).
At the highest redshifts we find that very few of the galaxies 
classified by eye as an early-type are within this region, and
in some cases these early-type classified galaxies are scattered
throughout the merger selection.  However, it is generally the case
that the galaxies picked out by eye as peculiars are within the
merger region and have more extreme Gini and \m20 values than
the early-types or disk galaxies.

\begin{figure*}
 \vbox to 140mm{
\includegraphics[angle=0, width=144mm]{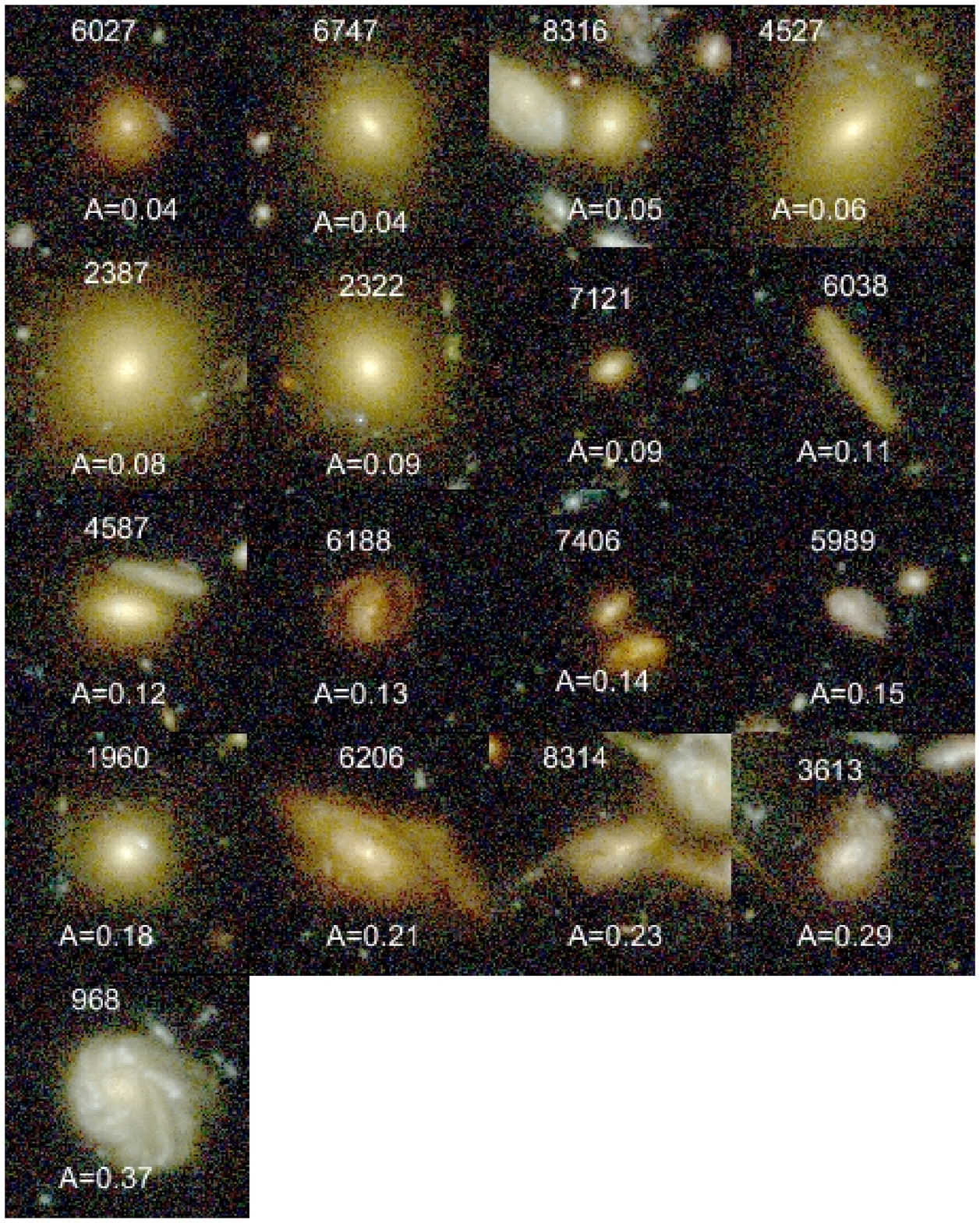}
 \caption{Colour images of galaxies with stellar masses M$_{*} >$ \lmass 
at redshifts $0.5 < z < 1.2$ in the Hubble Ultra Deep Field.  Plotted on
the top of each images is the ID number from Coe et al. (2006), and the bottom
number is the computed value of the asymmetry index in the B-band ($A_{\rm B}$). }
} \label{sample-figure}
\vspace{5cm}
\end{figure*}

\subsection{Merger Comparisons}

In this section we examine how the various methods used to define
merging galaxies compare with each other.  We limit this discussion
to the merging galaxies, as identifying other galaxy types, such as
ellipticals and especially different types of disk galaxies is
more difficult and will likely require additional indices 
(e.g., Conselice 2003).

There have been various propositions for how to identify mergers
within the CAS and Gini/\m20 space. As described in Conselice (2003)
one method for finding mergers is to use the conditions,

\begin{equation}
A > 0.35\, \&\, A > S,
\end{equation}

\noindent where this criteria is basically that the asymmetry must be
higher than an absolute limit of 0.35, which is the limit in which
local mergers are found (Conselice et al. 2000a,b; Conselice 2003),
and the relative definition of $A > S$ ensures that the asymmetric
light is not dominated by clumpy star-forming regions.

Equation (5) is a strong restriction, and will likely miss
many galaxies that are within some phase of a merger, as is
already known from examining a large sample of nearby galaxy
mergers (Conselice 2003).   This was
later calibrated by Conselice (2006) using N-body models
of galaxy mergers, where it was found that this condition will
identify a merger during 0.3-0.5 Gyr of its evolution. Thus,
during the merger process the asymmetry index will only find
a major merger in progress for part of the merger, and will not
be sensitive to minor mergers which have mass ratios of 1:5 and
greater (Conselice 2006). 

On the other hand, mergers in Gini/M20 are defined 
by Lotz et al. (2006) to be systems which meet the criteria,

\begin{equation}
G > -0.115 \times M_{20} + 0.384.
\end{equation}

\begin{figure*}
 \vbox to 140mm{
\includegraphics[angle=0, width=144mm]{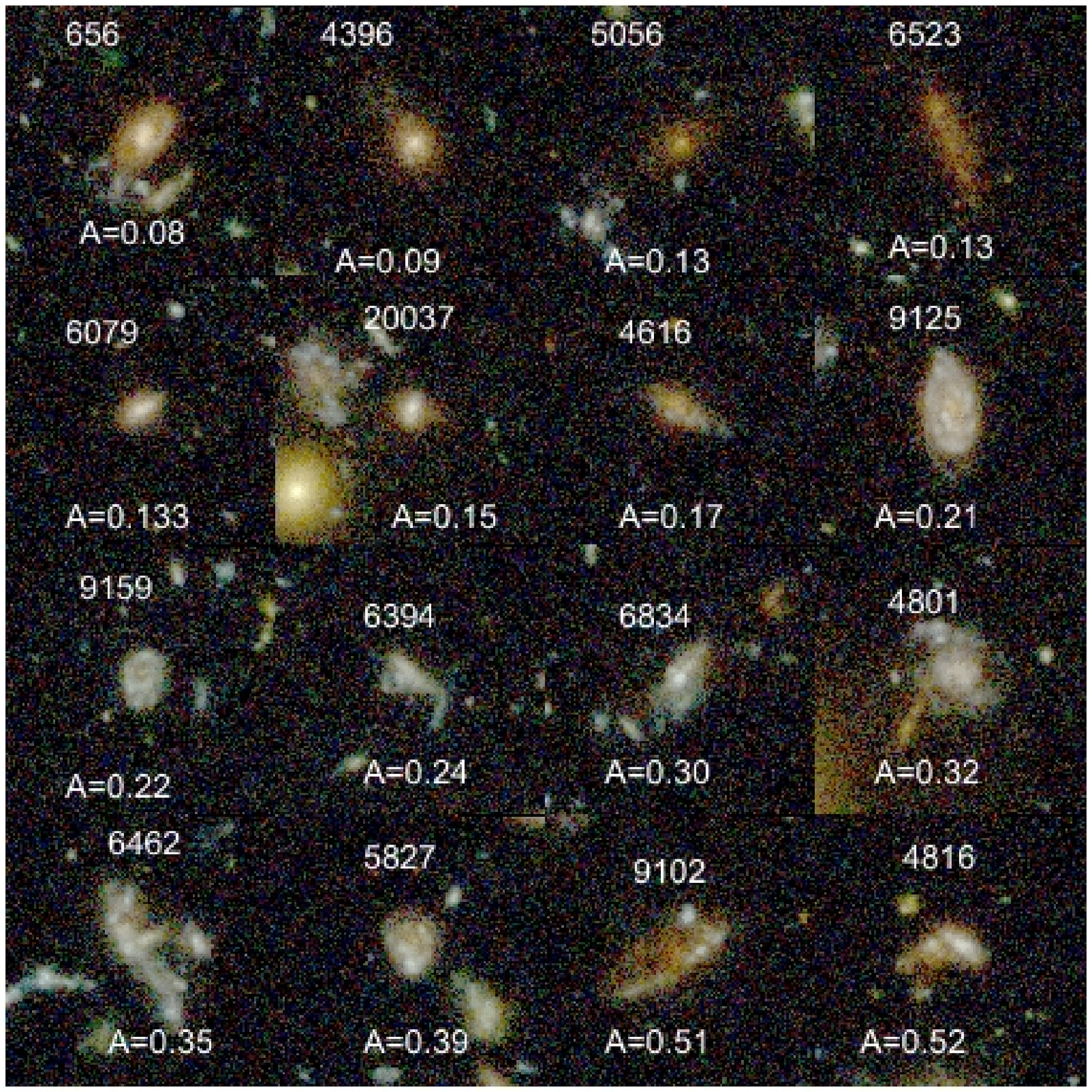}
 \caption{Galaxies with stellar masses M$_{*} >$ \lmass at redshifts
$1.2 < z < 1.6$.  Plotted on
the top of each images is the ID number from Coe et al. (2006), and the bottom
number is the computed value of the asymmetry index in the B-band ($A_{\rm B}$).}
} \label{sample-figure}
\vspace{2cm}
\end{figure*}

\noindent The first test of the merger method is to see how galaxies identified
within the CAS and Gini/\m20 methods compare. First, we show in Figure~8 
where galaxies
identified as mergers with the CAS merger criteria (eq. 5) would fall on the
Gini-\m20 digram.  We also show on Figure~9 all 
galaxies above the Gini/\m20 limit for mergers (eq. 6) re-plotted on the 
concentration vs. asymmetry diagram.  We do this to determine if the galaxies 
which would be identified as a merger within the Gini/\m20 space are the same 
as those in the merger category in CAS space, and vice-versa.
We can conclude from these figures that  almost all galaxies above the 
CAS limit would be identified as a merger within the Gini/\m20 plane 
(Figure~8), but
the reverse is not necessarily true.

As Figure~9 shows, almost all galaxies identified as
non-mergers in the Gini/\m20 space are located in the normal galaxy
region of CAS space.  However, when we examine the location of galaxies
identified as mergers within the Gini/\m20 space, and replot them on
the CAS space, we find that these galaxies occupy a large volume of this
space.  There are many systems which fall into the $A > 0.35$ region,
but the majority of galaxies are within the region of CAS space where
normal galaxies are expected.  What this means is that either
the Gini/\m20 criteria is not locating true mergers, or that it is 
much more sensitive to various merger phases and time-scales than
the CAS indices.

\section{The Merger History up to $z \sim 3$}

\subsection{Redshift and Morphological K-Corrections}

\subsubsection{Signal to Noise Reduction in CAS parameters}

A very important issue that we must address in this study, which compares 
properties of galaxies at different redshifts, is the fact that 
measured parameters, as well as the detection of galaxies themselves, 
can change solely due to the result of redshift and distant effects. 
The rapidly increasing luminosity distance of galaxies, with the 
correspondingly slowly changing angular size distance, 
produces a $(1 + z)^{4}$ decline in surface brightness.  Although
within the UDF we are easily detecting all galaxies down to our mass
limits (Conselice et al. 2003a), changes in the 
signal-to-noise ratio (S/N) and resolution due to redshift can mask, or 
mimic, real evolution in structural parameters (Conselice 2003; Conselice
et al. 2003a). We address these issues using simulations and apply 
this information to correct our asymmetry measurements, and
constrain possible evolution.

Two types of simulations were performed, those on nearby galaxies placed
at high redshifts (Conselice 2003), and by simulating galaxies imaged
at $z \sim 0.5$ placed at higher redshifts (see also Conselice et al. 2003a).
First, we carried out simulations to determine how galaxy structures 
change by using galaxies within the lowest
redshift bin around $z \sim 0.5$, and which have M$_{\rm B} < -18$,
roughly the limit of our mass sample with M$_{*} > 10^{8}$ \solm.  

The images of these objects in their rest-frame B-band morphology are 
simulated as they would appear at various redshifts from z = 1 to 3.
These simulations were done by creating a new background for these
galaxies with the same noise characteristics as the real data, then randomly 
placing the simulated galaxies into these backgrounds. The galaxies are 
reduced in resolution, signal to noise, flux, and surface brightness, and 
convolved with the PSF (see also Conselice et al. 2003a).

\begin{figure*}
 \vbox to 140mm{
\includegraphics[angle=0, width=144mm]{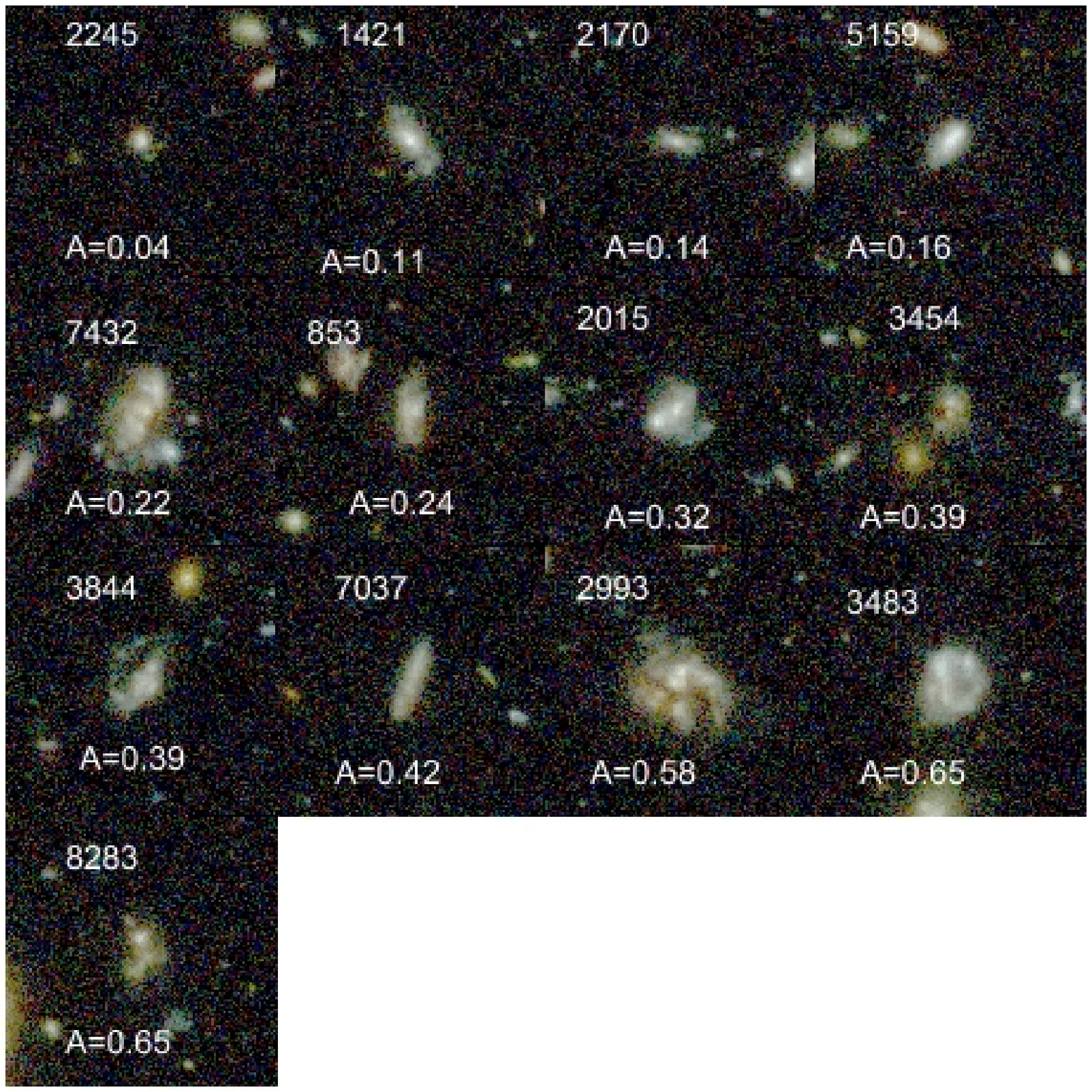}
 \caption{Galaxies with stellar masses M$_{*} >$ \lmass at redshifts
$1.6 < z < 2.2$.  Plotted on
the top of each images is the ID number from Coe et al. (2006), and the bottom
number is the computed value of the asymmetry index in the B-band ($A_{\rm B}$).}
} \label{sample-figure}
\vspace{2cm}
\end{figure*}

The result of all of these simulations is that the measured values of the 
asymmetries, and the other CAS parameters, become lower 
at higher redshifts (see also Conselice et al. 2003a; Conselice 2003). 
The average corrections necessary to account 
for these effects are generally low, with 
differences $\delta A = 0.04 - 0.06$ for the redshift ranges studied 
in this paper.

Furthermore, we find a very slight difference in retrieved asymmetries for 
fainter vs. brighter galaxies with different magnitudes in each redshift 
range. We find that the fainter galaxies are generally affected by noise 
more than brighter ones, and hence the systematics effects (and corrections) 
are larger, by roughly $\delta A \sim 0.02$, on average. As this is 
usually smaller than the random measurement errors for these faint galaxies, 
we do not account for this small difference. As discussed in Conselice
et al. (2000a), where similar simulations are done in terms of absolute 
S/N and resolution, the asymmetry index is not greatly affected by the 
reduced resolution and lower S/N for $z < 3$ galaxies with M$_{\rm B} < 
-18$ within the UDF.

\begin{figure*}
 \vbox to 140mm{
\includegraphics[angle=0, width=144mm]{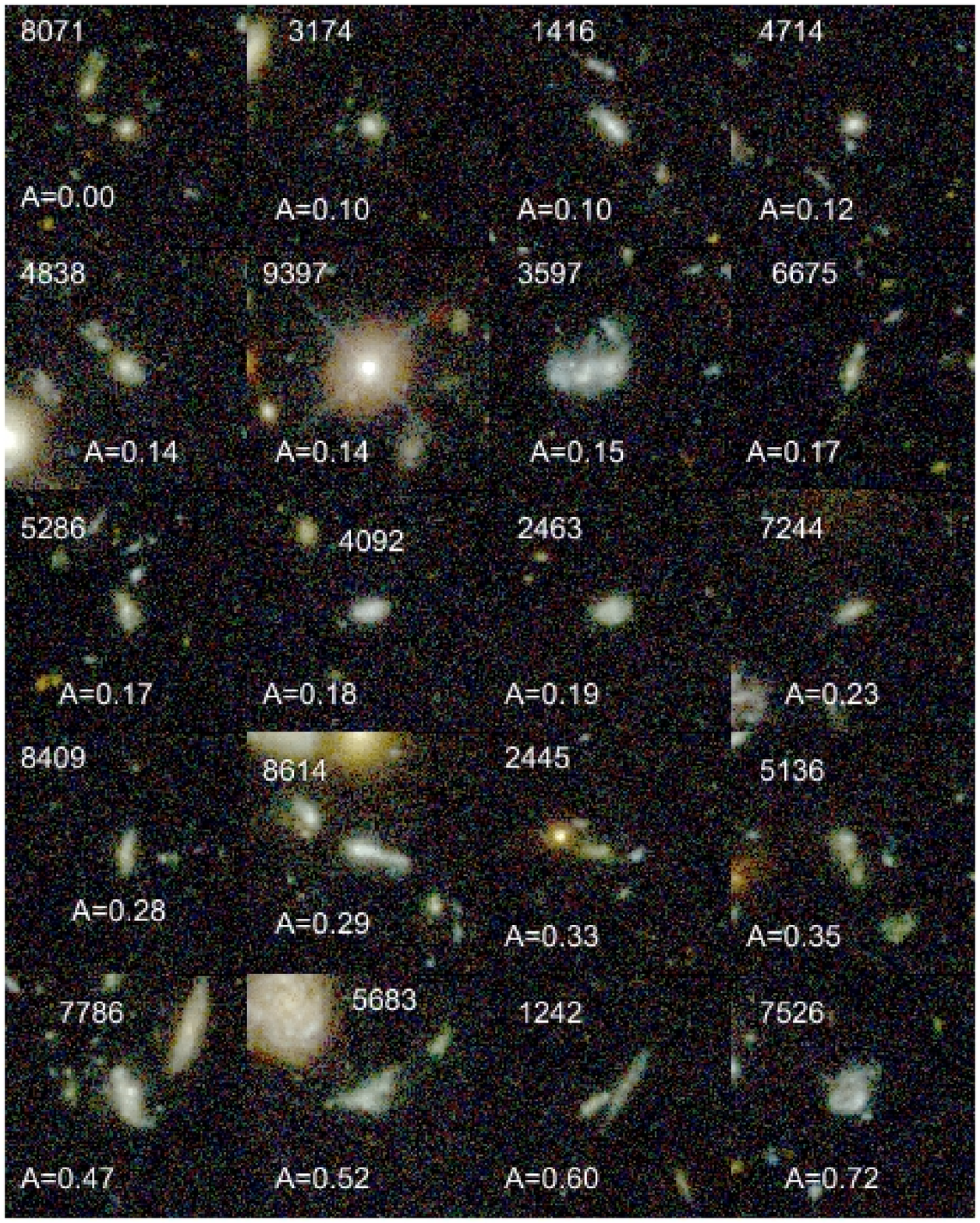}
 \caption{Galaxies with stellar masses M$_{*} >$ \lmass at redshifts
$2.2 < z < 3$.  Plotted on
the top of each images is the ID number from Coe et al. (2006), and the bottom
number is the computed value of the asymmetry index in the B-band ($A_{\rm B}$).}
} \label{sample-figure}
\vspace{5cm}
\end{figure*}

\subsubsection{Quantitative Morphological K-Correction}

The above corrections are however for galaxies affected just by signal
to noise and resolution due to redshift, assuming that the structure
of the galaxy is inherently the same. However, in our study we
are examining galaxies in the observed $z_{850}$ band, which corresponds
to various rest-frame wavelengths at different redshifts (Figure~3).  
This results in the observed $z_{850}$-band probing bluer light at
higher redshifts.  

We can account for this change using the results from \S 4.2, where we
calculate changes in the CAS and Gini/\m20 parameters at $z < 1.25$. At
these redshifts we are probing with the ACS filters the rest-frame 
optical and rest-frame UV for the same systems, allowing us to compare 
how indices change at shorter wavelengths probed with ACS at 
higher redshift.  Table~1
tabulates the change in the CAS Gini/\m20 parameters as a function of
redshift.  Using this table, we can calculate the average off-set between
the rest-frame observed values and those at rest-frame B-band ($\sim$ 
0.45 $\mu$m).  This change in parameters with wavelength is a function of 
galaxy type.  Thus, we can correct the CAS values to their rest-frame B-band
values depending on their eye-ball estimates of  morphological type.

\begin{figure*}
 \vbox to 150mm{
\includegraphics[angle=0, width=184mm]{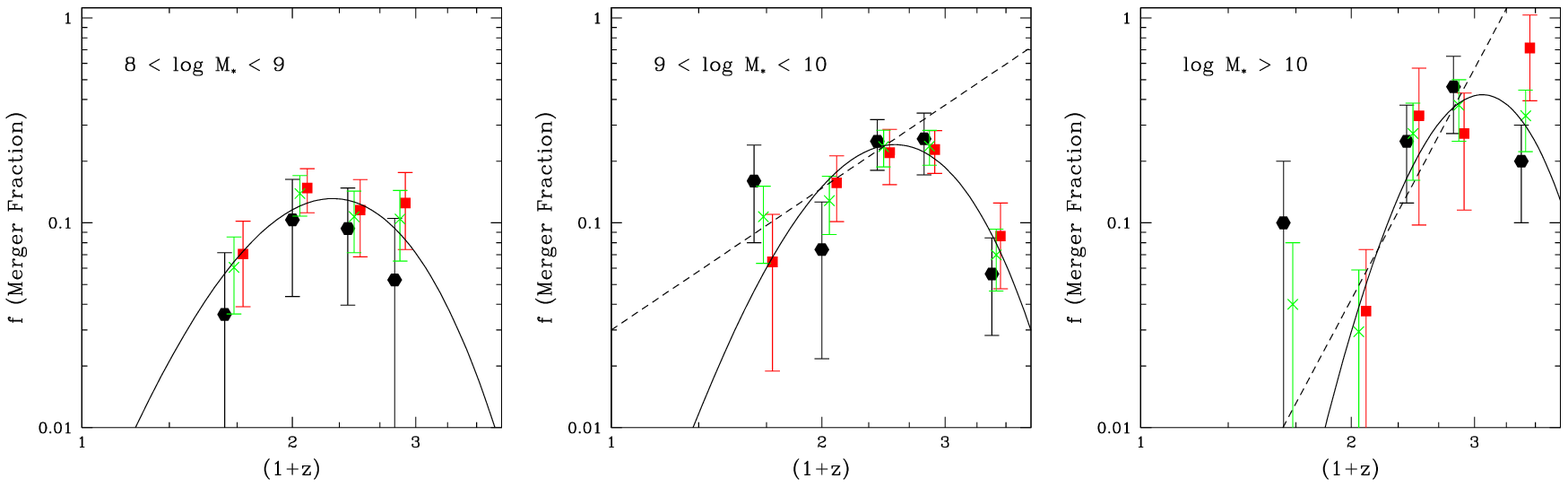}
 \caption{The merger fraction as a function of redshift and stellar
mass. Shown are the merger fractions at mass limits of $8 <$ log M$_{*} < 9$,
$9 <$ log M$_{*} < 10$, and log M$_{*} >$ 10.  The black circles show
the merger fraction as measured in the UDF, while the red boxes are
the merger fractions from the HDF-North where we can probe the 
rest-frame optical (Conselice et al. 2003a).  The green crosses show
the evolution of the merger fraction for a combined UDF and HDF-N sample.}
} \label{sample-figure}
\vspace{-7cm}
\end{figure*}

For example, at $z \sim 2.5$ the rest-frame wavelength probed by the 
$z_{850}$-band 
is roughly $\lambda_{\rm z-rest} = 0.32\,\mu$m.  Using Table~1 this gives us
a changed of $\Delta A$ = $-0.10$ for Peculiars, $-0.06$ for spirals and 
$-0.03$ for
ellipticals, using the $z = 0.75 - 1.25$ results. These changes are
slightly different using the $z = 0.25 - 0.75$ values, although the
general overall reduction in asymmetry is similar.  We apply this 
correction to each galaxy, depending on its wavelength and morphological
type when we compute the merger fraction.  

If we denote this morphological $k-$correction as $\Delta A_{\rm k-corr}$,
and the redshift correction (\S 5.1.1) as $\Delta A_{\rm z}$, then
the final asymmetry we use for our measurements of the merger fraction is
given by

\begin{equation}
A_{\rm B,final} = A(z_{850},z,T) + \Delta A_{\rm k-corr} + \Delta A_{\rm z},
\end{equation}

\noindent where the $A_{\rm B,final}$ is the derived rest-frame B-band
morphology of the galaxy under study, and $A(z_{850},z,T)$ is the observed
asymmetry in the $z_{850}$-band, which is a function of the redshift $z$,
and morphological-type ($T$).  Note that in our case, $\Delta A_{\rm k-corr}
< 0$ and $\Delta A_{\rm z} > 0$ at $z > 1.5$.

We use these corrections  to utilise the CAS definition for finding
major-mergers, which requires that $A_{\rm B} > 0.35$.  As explained in
detail in Conselice (2006), there are major-mergers with $A_{\rm B} < 0.35$,
and thus we are not complete in finding mergers by using this definition.
However, because this asymmetry limit is well defined, it allows us to
determine the time-scale for merging and thus convert the merger fraction
measured with the $A_{\rm B} > 0.35$ limit into a merger rate.

\subsection{Visual Appearance of Massive Galaxies}

Before we determine the merger history for our sample of galaxies, it is
instructive to examine the visual structures of our sample and the
corresponding asymmetries of these galaxies.  As an example, we examine
how the most massive galaxies look as a function of redshift.
Figures~10 through 13 show the visual morphologies of
galaxies with M$_{*} >$ \lmass up to redshifts $z \sim 3$.  We plot
the final  asymmetry values for these galaxies in each of their respective 
panels.
There are a few interesting trends in the apparent structures of
these galaxies. The first is that in the lowest redshift bin, $0.5 < z < 1.2$
a large fraction of the M$_{*} >$  \lmass galaxies have a regular normal
elliptical or spheroid-like appearance.  This is confirmed by where these
galaxies fall in the CAS and Gini/\m20 space (e.g., Figure 5; see also
Conselice et al. 2007b).  As can also be seen, these massive galaxies
are generally reddish in appearance, which is also the result when
examining galaxy colour and stellar mass (e.g., Bundy et al. 2006).

\begin{figure*}
 \vbox to 150mm{
\includegraphics[angle=0, width=184mm]{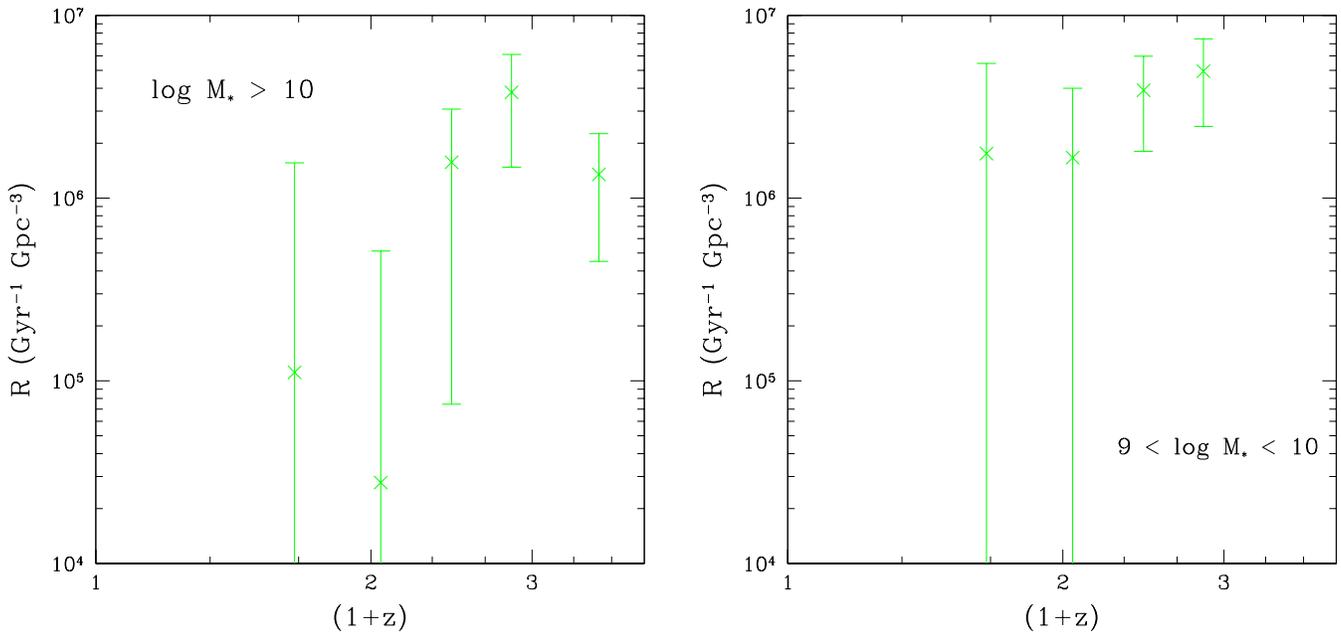}
 \caption{The merger rate in units of Gyr$^{-1}$ Gpc$^{-3}$ for
the combined UDF and HDF-N samples.  The merger rate is given 
by equation (10) where we include number densities as measured
by Drory et al. (2005) and time-scales for CAS mergers
from Conselice (2006).}
} \label{sample-figure}

\vspace{-5cm}
\end{figure*}

A few of the galaxies within the range $0.5 < z < 1.2$ appear to be peculiars
or spiral galaxies.  This is even more the case when we examine 
massive galaxies at higher redshifts (Figure~11-13).  This change can
been seen even more significantly in the $1.2 < z < 1.6$ redshift range.
While there are some galaxies at stellar masses M$_{*} >$ \lmass which 
appear to
be roughly spherical, and in some cases red (e.g., object 5056, Figure~11),
but already
the majority of these objects have an appearance which suggests that these
galaxies are undergoing some kind of formation either through star formation
in a disk or through a merger process of some type.   In this redshift
range there are a few obvious examples of galaxies which are undergoing
some type of merger and have resulting high asymmetries (objects 6462, 9102 
and 4816).

In the highest redshift ranges $1.6 < z < 2.2$ (Figure~12) 
and $2.2 < z < 3.0$ (Figure~13) there are more irregular/merging galaxies.  
There are however interesting exceptions to this. In Figure~12 and 13 there 
are cases of galaxies which look normal, and are perhaps 
morphological progenitors of elliptical galaxies (e.g., objects 2245, 8071, 
3174).   There are also examples of galaxies at these redshift ranges which 
have a disk
like appearance (e.g., objects 1421, 2170, and 5159).  Clearly, as discussed 
within the Hubble Deep Field, there is a diversity of morphologies for the
most massive galaxies even at $z > 2$ (Conselice et al. 2003a).  This might 
be the result of these galaxies undergoing an inhomogeneous formation history 
with some galaxies undergoing merging, and others in a more quiescent 
state.  In the next section we investigate merger fractions based on 
these structural appearances and their measured CAS parameters.

\subsection{Merger Fractions}

One of the major questions in extragalactic studies is the
role of mergers in the formation of galaxies. There
are  various types of mergers - minor and
major, and the relative role of these over cosmic time is largely
unknown.
One of the great benefits of using the CAS system for
finding mergers is that it allows us to quantify the
merger fraction, merger rates, and thus the number of 
mergers occurring in a galaxy population 
(Conselice et al. 2003a; Conselice 2006).  While in this
section we investigate the merger fractions based on
standard techniques we have developed (e.g., Conselice
2003; Conselice 2006; Bridge et al. 2007), we examine
in detail those galaxies that we consider mergers later in a future
paper in this series.

\begin{figure}
 \vbox to 120mm{
\includegraphics[angle=0, width=90mm]{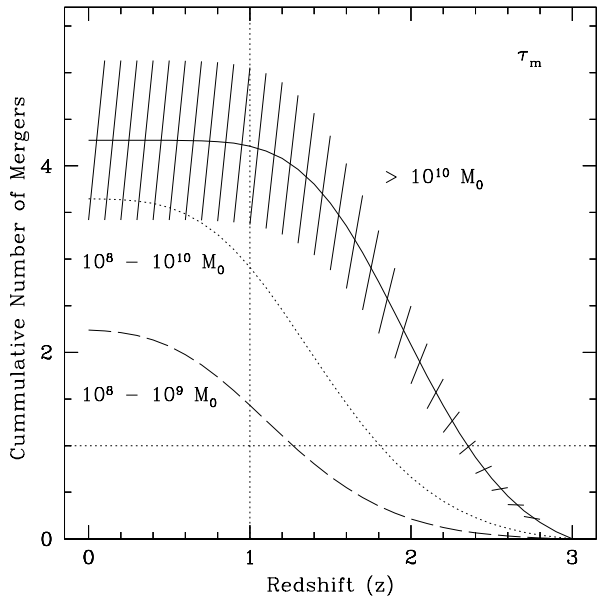}
 \caption{The cumulative number of major mergers for galaxies of
various masses as a function of redshift. These cumulative mergers
beginning at $z = 3$. The error range for the M$_{*} >$ \lmass
galaxies is shown.  In total, the most massive galaxies have
a larger number of total galaxy mergers than the lower mass systems.}
} \label{sample-figure}
\end{figure}

The first observation we can derive from our
CAS values is the evolution of the merger fraction. We determine the
merger fraction for galaxies of various masses using
the criteria from equation (5).   There are a few caveats
to measuring the merger fraction which we must consider
before using these values to determine how galaxies
are evolving due to major mergers, or even to measure
the merger fraction. Our final merger fraction 
values are tabulated in Table~2.

\vspace{1cm}
\setcounter{table}{1}
\begin{table*}
 \caption{Merger fractions as a function of stellar mass and redshift}
 \label{tab1}
 \begin{tabular}{@{}cccc}
  \hline
\hline
z (UDF) & f(10$^{8}$ \solm $- 10^{9}$ \solm) & f(10$^{9}$ \solm $- 10^{10}$ \solm) & f($> 10^{10}$ \solm) \\
\hline
0.6 & 0.04$\pm$0.04 & 0.16$\pm$0.08 & 0.10$\pm$0.10 \\
1.0 & 0.10$\pm$0.06 & 0.07$\pm$0.05 & 0.00$\pm$0.00 \\
1.4 & 0.09$\pm$0.05 & 0.25$\pm$0.07 & 0.25$\pm$0.13 \\
1.8 & 0.06$\pm$0.05 & 0.26$\pm$0.09 & 0.46$\pm$0.19 \\
2.5 & ...      & 0.06$\pm$0.03 & 0.20$\pm$0.10 \\
\hline
z (HDF) & f(10$^{8}$ \solm $- 10^{9}$ \solm) & f(10$^{9}$ \solm $- 10^{10}$ \solm) & f($> 10^{10}$ \solm) \\
\hline
0.6 & 0.07$\pm$0.03 & 0.06$\pm$0.05 & 0.00$\pm$0.00 \\
1.0 & 0.15$\pm$0.04 & 0.16$\pm$0.06 & 0.04$\pm$0.04 \\
1.4 & 0.12$\pm$0.05 & 0.22$\pm$0.07 & 0.33$\pm$0.24 \\
1.8 & 0.13$\pm$0.05 & 0.23$\pm$0.05 & 0.27$\pm$0.16 \\
2.5 & ...       & 0.09$\pm$0.04 & 0.71$\pm$0.32 \\
\hline
z (Combined) & f(10$^{8}$ \solm $- 10^{9}$ \solm) & f(10$^{9}$ \solm $- 10^{10}$ \solm) & f($ > 10^{10}$ \solm) \\
\hline
0.6 & 0.06$\pm$0.02 & 0.11$\pm$0.04 & 0.04$\pm$0.04 \\
1.0 & 0.14$\pm$0.03 & 0.13$\pm$0.04 & 0.03$\pm$0.03 \\
1.4 & 0.11$\pm$0.04 & 0.24$\pm$0.05 & 0.27$\pm$0.11 \\
1.8 & 0.10$\pm$0.04 & 0.24$\pm$0.05 & 0.38$\pm$0.13 \\
2.5 & ...       & 0.07$\pm$0.02 & 0.33$\pm$0.11 \\
\hline
\end{tabular}
\end{table*}

First, it is important to note that the merger fraction we
measure is in no sense the total
galaxy merger fraction. 
All merger fractions we derive are computed using a certain technique,
in this case the CAS parameters,
and this technique is only sensitive to a well defined
time-range. In the case of CAS mergers, this time-span is
roughly 0.4 Gyr (Conselice 2006). As determine in Conselice
(2006) there are phases of a merger which will not be picked
up by the CAS technique.  A different technique will find
a different merger fraction if it has a time sensitivity different
from the CAS system. For example galaxy pair methods have
roughly a factor of two different time-scale for a merger
than the CAS system (De Propris et al. 2007), but gives
the same merger rate.  We therefore expect galaxies that by eye 
appear as a merger, but will not have a high asymmetry.  

There are a few ways to fit the merger fraction. The first
is the traditional power-law format (Patton et al. 2002; Conselice
et al. 2003; Bridge et al. 2007).  This fitting format is given
by,

\begin{equation}
f_{\rm m}(z) = f_{0} \times (1+z)^{m}
\end{equation}

\noindent where $f_{\rm m}(z)$ is the merger fraction at a given
redshift, $f_{0}$ is the merger fraction at $z = 0$, and
$m$ is the power-law index for characterising the merger
fraction evolution.

Another way to characterise the merger fraction evolution which
dates back to theoretical arguments based on Press-Schechter formalism
for merging (Carlberg
1990) is a combined power-law exponential evolution.  This form
appears to be a better fit to all of the redshift data than a simple
power-law (Conselice 2006).  The formula for this evolution is given by:

\begin{equation}
f_{\rm m} = \alpha (1+z)^{m} \times {\rm exp}(\beta(1+z)),
\end{equation}

\noindent where the $z = 0$ merger fraction is given by
$f_{\rm m}(0) = \alpha \times$ exp($\beta$).

We plot in Figure~14 the merger fraction for the sample of galaxies
in the UDF as well as the HDF-N, where there is NICMOS high-resolution
imaging of galaxies, and thus the merger fraction can be measure
strictly in the rest-frame optical at all redshifts.  The results derived
within both fields are however very similar.  We show
the combined HDF-N and UDF merger fractions as the green crosses on
Figure~14.  The merger fractions for the UDF are always measured
in the observed $z_{850}$-band.  We however correct for both the
effects of redshift, as well as for the morphological $k-$correction
using the rest-frame B-band asymmetry calculated using equation (7).

We show as the solid
line on Figure~14 the best fit values for the merger fraction using
the combined exponential/power-law fitting formula (eq. 9). We also 
fit up to the merger fraction redshift
peak the power-law form given by equation (8) for all but
the \lllmass $<$ M$_{*} <$ \llmass galaxies where there
are too few points to carry out a reliable fit.  The values
found for the M$_{*} >$ \lmass galaxies are: $\alpha = 0.01\pm0.02$,
$m = 33\pm16$, and $\beta = -10\pm5$ for the exponential/power-law
formalism.   The lower mass
systems have a systematically higher value of $\alpha$, a lower value of
$m$, and a larger (less negative) value of $\beta$.




The peak merger fraction within the exponential/power-law fit 
occurs at a redshift of $z_{\rm peak} = -(1+m/\beta)$.
Using our fits, we calculate that the peak in the merger fraction
occurs at redshifts $z_{\rm peak} = 2.08$ for galaxies
with M$_{*} >$ \lmass, and $z_{\rm peak} = 1.54$, and $z_{\rm peak} = 1.28$ for
galaxies with \llmass $<$ M$_{*} <$ \lmass and \lllmass $<$ M$_{*} <$ \llmass
selected galaxies, respectively.  This suggests that the peak merger fraction
appears to occur later for lower-mass galaxies. The downsizing for the 
lower-mass galaxies therefore might occur because these systems
are not merging as quickly as the higher mass galaxies.

\subsection{Merger Rates and the Cumulative Number of Mergers}

\subsubsection{The Galaxy Merger Rate at $z < 3$}

By using the number densities for galaxies within our mass ranges
from previous work, and the
time-scales from our CAS method, we can calculate
the merger rate for our M$_{*} >$ \lllmass galaxy samples.
The number densities for our systems are taken from
Drory et al. (2005), and the time-scales 
for merging are derived from equation 10 in Conselice (2006), 
based on N-body models analysed using CAS indices. 

The galaxy merger rate for our systems can be calculated through
the merger rate equation,

\begin{equation}
\Re(z) = {\rm f_{gm}}(z) \cdot \tau_{\rm m}^{-1} {\rm n_{gm}}(z)
\end{equation}

\noindent where n$_{\rm gm}$ is the number densities of
galaxies within a given stellar mass range, and f$_{\rm gm}$ is the 
galaxy merger fraction. Note that
this is not the merger fraction, which is
the number of mergers divided by the number of galaxies, which is roughly 
half the galaxy merger fraction (Conselice 2006).   We convert our
merger fractions used earlier in this paper to galaxy merger fractions
using the relation,

\begin{equation}
f_{gm} = \frac{2 \times f_{m}}{1+1/f_{m}}.
\end{equation}

\noindent The time-scales we use in eq. (10) ($\tau_{\rm m}$) come from 
the calculations based on N-body models
in Conselice (2006). There is a slight decrease in the time-scale
for galaxies with lower masses. We utilise these results to calculate that
the merger time-scale for the most massive galaxies with M$_{*} >$ \lmass
is $\tau_{\rm m} = 0.34$ Gyr, and slightly lower, between 0.27-0.29 Gyr
for the lower mass systems.  

By utilising the number densities for galaxies of a given stellar mass
from Drory et al. (2005), the time-scales from Conselice (2006), and the
galaxy merger fractions from this paper, we compute the merger
rate as a function of stellar mass using equation (10).  The results of
this calculation are shown in Figure~15.  As can be seen, the merger rate
decreases at lower redshifts for galaxies at masses \llmass
$<$ M$_{*}$ $<$ \lmass, with the average merger rate, $<\Re>$ 
$\sim 3 \times 10^{6}$ galaxies merging Gyr$^{-1}$ Gpc$^{-3}$.

The more massive galaxies with M$_{*} >$ \lmass have a lower
merger rate at all redshifts, and a more steeply declining
merger rate with time.  We find that the merger rate 
declines as $\sim (1+z)^{5.5\pm2.5}$ for these massive galaxies, 
which is similar to the decline in the merger fraction.  The
merger rate for \llmass $<$ M$_{*}$ $<$ \lmass galaxies evolves
as $\sim (1+z)^{2.6\pm0.6}$, again similar to the evolution in
the merger fraction, and showing a more gradual decline with
time compared to more massive galaxies.  The steeper decline in the 
merger rate for more
massive galaxies is likely partially the reason the star
formation rate in these systems also drops much faster
than for lower mass systems (Conselice et al. 2007b).







\subsubsection{Total Number of Mergers at $z < 3$}


One of the major benefits to calculating the merger rate
is that it allows us to determine the total number of major
mergers a galaxy of a given mass will undergo between two
redshifts.  We calculate the total number of major
mergers a galaxy with M$_{*} >$ \lmass undergoes
from $z \sim 3$ to $z \sim 0$ using equation (11)
in Conselice (2006).  We use our fitted exponential/power-law
functions for this calculation, which has an associated
uncertainty associated with with the fit.

Figure~16 shows the cumulative number of mergers 
which have occurred from $z \sim 3$ to $z \sim 0$ as a function
of mass. Although
we do not have a good idea of the merger fraction or rate for
our massive galaxies at $z < 0.6$, we know from previous work that
the merger fractions for these systems is very low, particularly
for the most massive galaxies (e.g.,
Patton et al. 2002; De Propris et al. 2007; 
Conselice et al. 2007b). Using our merger fraction
fits, we calculate that the average
number of mergers a  galaxy with M$_{*} >$ \lmass
will undergo from $z \sim 3$ to 0 is N$_{\rm m} = 
4.3^{+0.8}_{-0.8}$.  This is nearly the same as the value
we obtained by using the HDF-N, where for the same
mass range, we found N$_{\rm m} = 4.4^{+1.1}_{-0.9}$
(Conselice 2006).  The uncertainties in this calculation
are dominated by those from the merger-time scale calculation.

The lower mass galaxies generally have a lower total number of major
mergers occurring at $z < 3$ (Figure~16).  This is perhaps partially
because we cannot measure the merger fraction for these lower mass
galaxies reliably at $z < 0.5$. The volume probed by the UDF
is not large enough to reliably trace this merger fraction, and other
fields which are large enough, are not deep enough.
In general however, it appears that the lower mass galaxies do not
undergo the same number of major mergers as the highest mass galaxies.  
Furthermore, it appears, as we discussed in Conselice (2006), that
most of the major mergers for the most massive galaxies occurs
at higher redshifts, $z > 1$, although for the lower mass galaxies
there is still some merging at these later times.  This is consistent
with the downsizing of star formation (e.g., Bundy et al. 2006) produced 
through merging, and the fact that 
peculiar galaxies found at
lower redshifts often contain a lower stellar mass (e.g., Bundy
et al. 2005).

\section{Summary}

This paper begins a series in which we examine the structures of
distant galaxies, and determine the likely role of galaxy merging
in the formation of galaxies. In this paper we examine structural
parameters using the CAS and Gini/\m20 system on galaxies found within
the Hubble Ultra Deep Field (UDF).  We also examine the stellar masses
and eye-ball estimates of morphological types for these systems. Our
major conclusions and findings include:

\noindent I. Down to $z_{850} = 27$ the majority of galaxies found in the
UDF are peculiar in appearance.  This suggests that galaxy
formation is actively ongoing for these systems.  

\noindent II. We compare how visual estimates of morphology agree
with positions in CAS and Gini/\m20 space. We find a generally good
agreement between these methods with the merger region in CAS
space nearly totally occupied by galaxies which  are visually
classified as peculiars/mergers.

\noindent III. We examine the merger fraction of galaxies in the UDF
using the CAS system and compare with our previous results of the merger
fraction and rate derived using the Hubble Deep Field. We confirm our 
earlier measurements of the merger history, including the fact that
the highest mass galaxies with M$_{*} >$ \lmass have a steeply
increasing merger fraction up to $z \sim 3$. This increase can be
fit as a power-law $\alpha (1+z)^{m}$, with $m \sim 6$ for the combined
UDF and HDF-N sample.  We find that the merger fraction for
lower mass galaxies are lower at high redshift, but reach
their peak fractions at lower redshifts. In total we
find that M$_{*} >$ \lmass galaxies undergo on average 4.3$^{+0.8}_{-0.8}$
major mergers at $0 < z < 3$, with most of this merging occurring at
$z > 1$.

We thank Dan Coe for making his analysis results of the UDF freely
available, and Kevin Bundy for assistance with the stellar masses.
We also acknowledge support from the University of Nottingham.

\appendix

\label{lastpage}


\begin{thebibliography}{99}
\bibitem[\protect\citeauthoryear{}{}]{b1} Abraham, R., van den Bergh, S., Nair, P. 2003, ApJ, 588, 218
\bibitem[\protect\citeauthoryear{}{}]{b1} Benitez, N. 2000, ApJ, 536, 571
\bibitem[\protect\citeauthoryear{}{}]{b1} Beckwith, S., et al. 2006, AJ, 132, 1729
\bibitem[\protect\citeauthoryear{}{}]{b1} Bershady, M.A., Jangren, J.A., Conselice, C.J. 2000, AJ, 119, 2645 
\bibitem[\protect\citeauthoryear{}{}]{b1} Bridge, C., et al. 2006, ApJ, 659, 931
\bibitem[\protect\citeauthoryear{}{}]{b1} Bruzual, G., Charlot, S. 2003, MNRAS, 344, 1000
\bibitem[\protect\citeauthoryear{}{}]{b1} Bundy, K., et al. 2006, ApJ, 651, 120
\bibitem[\protect\citeauthoryear{}{}]{b1} Bundy, K., Ellis, R.S., Conselice, C.J. 2005, ApJ, 625, 621
\bibitem[\protect\citeauthoryear{}{}]{b1} Carlberg, R. 1990, ApJ, 359, 1L
\bibitem[\protect\citeauthoryear{}{}]{b1} Cassata, P., et al. 2005, MNRAS, 357, 903
\bibitem[\protect\citeauthoryear{}{}]{b1} Coe, D., Benitez, N., Sanchez, S.F., Jee, M., Bouwens, R., Ford, H. 2006, AJ, 132, 926
\bibitem[\protect\citeauthoryear{}{}]{b1} Conselice, C.J. 1997, PASP, 109, 1251
\bibitem[\protect\citeauthoryear{}{}]{b1} Conselice, C.J., Bershady, M.A., Jangren, A. 2000a, ApJ, 529, 886
\bibitem[\protect\citeauthoryear{}{}]{b1} Conselice, C.J., Bershady, M.A., Gallagher, J.S. 2000b, A\&A, 354, 21L
\bibitem[\protect\citeauthoryear{}{}]{b1} Conselice, C.J., Gallagher, J.S., Calzetti, D., Homeier, N., Kinney, A. 2000c, AJ, 119, 79
\bibitem[\protect\citeauthoryear{}{}]{b1} Conselice, C.J. 2003, ApJS, 147, 1
\bibitem[\protect\citeauthoryear{}{}]{b1} Conselice, C.J., Bershady, M.A., Dickinson, M., Papovich, C. 2003a, AJ, 126, 1183
\bibitem[\protect\citeauthoryear{}{}]{b1} Conselice, C.J., Chapman, S.C., Windhorst, R.A. 2003b, ApJ, 596, 5L
\bibitem[\protect\citeauthoryear{}{}]{b1} Conselice, C.J., Gallagher, J.S., Wyse, R.F.G. 2002, AJ, 123, 2246
\bibitem[\protect\citeauthoryear{}{}]{b1} Conselice, C.J., et al. 2004, ApJ, 600, 139L
\bibitem[\protect\citeauthoryear{}{}]{b1} Conselice, C.J., Blackburne, J., Papovich, C. 2005a, ApJ, 620, 564
\bibitem[\protect\citeauthoryear{}{}]{b1} Conselice, C.J., Bundy, K., Ellis, R., Brichmann, J., Vogt, N., Phillips, A. 2005b, ApJ, 628, 160
\bibitem[\protect\citeauthoryear{}{}]{b1} Conselice, C.J. 2006a, MNRAS, 373, 1389
\bibitem[\protect\citeauthoryear{}{}]{b1} Conselice, C.J. 2006b, ApJ, 638, 686
\bibitem[\protect\citeauthoryear{}{}]{b1} Conselice, C.J., et al. 2007a, ApJ, 660, 55L
\bibitem[\protect\citeauthoryear{}{}]{b1} Conselice, C.J., et al. 2007b, MNRAS, in press, arXiv:0708.1040
\bibitem[\protect\citeauthoryear{}{}]{b1} Courteau, S., McDonald, M., Widrow, L.M., Holtzman, J. 2007, ApJ, 655, 21
\bibitem[\protect\citeauthoryear{}{}]{b1} De Propris, R., Conselice, C.J., Driver, S.P., Liske, J., Pattton, D., Graham, A., Allen, P. 2007, preprint, arXiv:0705.2528
\bibitem[\protect\citeauthoryear{}{}]{b1} Dickinson, M., et al. 2000, ApJ, 531, 624
\bibitem[\protect\citeauthoryear{}{}]{b1} Driver, S., et al. 1998, ApJ, 496, 93L
\bibitem[\protect\citeauthoryear{}{}]{b1} Drory, N., et al. 2005, ApJ, 619, 131L
\bibitem[\protect\citeauthoryear{}{}]{b1} Elmegreen, D.M., Elmegreen, B.G., Rubin, D.S., Schaffer, M.A. 2005, ApJ, 631, 85
\bibitem[\protect\citeauthoryear{}{}]{b1} Faber, S.M., et al. 2007, ApJ, 665, 265
\bibitem[\protect\citeauthoryear{}{}]{b1} Giavalisco, M., et al. 2004, ApJ, 600, 93L
\bibitem[\protect\citeauthoryear{}{}]{b1} Glazebrook, K., Ellis, R., Santiago, B., Griffiths, R. 1995, MNRAS, 275, 19L
\bibitem[\protect\citeauthoryear{}{}]{b1} Grogin, N.A., et al. 2005, ApJ, 627, 97L
\bibitem[\protect\citeauthoryear{}{}]{b1} Hernandez-Toledo, H.M., Avila-Reese, V., Conselice, C.J., Puerari, I. AJ, 2005, AJ, 129, 682
\bibitem[\protect\citeauthoryear{}{}]{b1} Hibbard, J.E., \& Vacca, W.D. 1997, AJ< 114, 1741
\bibitem[\protect\citeauthoryear{}{}]{b1} Lotz, J.M., Primack, J., Madau, P. 2004, AJ, 128, 163L
\bibitem[\protect\citeauthoryear{}{}]{b1} Lotz, J.M., et al. 2006, astro-ph/0602088
\bibitem[\protect\citeauthoryear{}{}]{b1} Moustakas, L.A., et al. 2004, ApJ, 600, 131L
\bibitem[\protect\citeauthoryear{}{}]{b1} Papovich, C., Dickinson, M., Giavalisco, M., Conselice, C.J., Ferguson, H.C. 2005, ApJ, 631, 101
\bibitem[\protect\citeauthoryear{}{}]{b1} Papovich, C., Giavalisco, M.,Dickinson, M. Conselice, C.J., Ferguson, H.C. 2003, ApJ, 598, 827
\bibitem[\protect\citeauthoryear{}{}]{b1} Patton, D.R., et al. 2002, ApJ, 565, 208
\bibitem[\protect\citeauthoryear{}{}]{b1} Ravindranath, S., et al. 2006, ApJ, 652, 963
\bibitem[\protect\citeauthoryear{}{}]{b1} Stanford, S.A., Dickinson, M.E., Postman, M., Ferguson, H.C., Lucas, R.A., Conselice, C.J., Budavari, T., Somerville, R. 2004, AJ, 127, 131
\bibitem[\protect\citeauthoryear{}{}]{b1} Taylor-Mager, V., Conselice, C., Windhorst, R., Jansen, R. 2007, ApJ, 659, 162
\bibitem[\protect\citeauthoryear{}{}]{b1} Teplitz, H.I., et al. 2006, AJ, 132, 853
\bibitem[\protect\citeauthoryear{}{}]{b1} Thompson, R.I., et al. 2005, AJ, 130, 1
\bibitem[\protect\citeauthoryear{}{}]{b1} Trujillo, I., Conselice, C.J., Bundy, K., Cooper, M.C., Eisenhardt, P., Ellis, R.S. 2007, MNRAS, in press, arXiv:0709.0621
\bibitem[\protect\citeauthoryear{}{}]{b1} Williams, R., et al. 1996, AJ, 112, 1335
\bibitem[\protect\citeauthoryear{}{}]{b1} Windhorst, R., et al. 2002, ApJS, 143, 113
\bibitem[\protect\citeauthoryear{}{}]{b1} Wright, S.A., et al. 2007, ApJ, 658, 78
\end{thebibliography}
\end{document}